\pdfoutput=1
\documentclass[twocolumn]{openjournal}

\usepackage{aas_macros}
\usepackage{graphicx}
\usepackage{dcolumn}
\usepackage{booktabs}
\usepackage{bm}
\usepackage{hyperref}
\usepackage{autobreak}
\usepackage{xcolor}
\usepackage{physics}
\usepackage{amsmath, amssymb} 
\usepackage{ragged2e}
\usepackage{soul} 
\usepackage{tikz,tkz-tab,amsmath}

\usepackage{tikz,lipsum,lmodern}
\usepackage[T1]{fontenc}
\usepackage[most]{tcolorbox} 
\usepackage{lmodern} 
\usepackage{float}
\usepackage{chemfig}  
\usepackage{wrapfig}
\usepackage{footmisc}

\def\aap{A\&A}                
\def\apj{ApJ}                 
\def\apjs{ApJS}                 
 
\def\prd{Phys.\ Rev.\ D}
\def\mnras{MNRAS}

\tcbset{colback=green!5,colframe=green!35!black,fonttitle=\bfseries}

\def\Om{\Omega_\mathrm{m}}
\def\fsky{f_{\text{sky}}}
\def\Msun{M_\odot}

\definecolor{RoyalBlue}{rgb}{0.25,.41,.88}
\definecolor{DeepPurple}{rgb}{.45,.1,.75}
\definecolor{WildStrawberry}{HTML}{EE2967}
\definecolor{RedWine}{rgb}{0.743,0,0}
\definecolor{BitterSweet}{rgb}{1.0, 0.44, 0.37}
\definecolor{BurntOrange}{rgb}{0.8, 0.33, 0.0}
\definecolor{MidnightGreen}{rgb}{0.0, 0.29, 0.33}

\begin{document}

\preprint{000-000-000}

\title{Late-time growth weakly affects the significance of high-redshift massive galaxies}

\author{Qianran Xia}
\email{qianranx@umich.edu}
\author{Dragan Huterer}%
\email{huterer@umich.edu}
\affiliation{Leinweber Center for Theoretical Physics, University of Michigan, 450 Church St, Ann Arbor, MI 48109-1040}
\affiliation{Department of Physics, College of Literature, Science and the Arts, University of Michigan, 450 Church St, Ann Arbor, MI 48109-1040}
\author{Nhat-Minh Nguyen}
\email{nhat.minh.nguyen@ipmu.jp}
\affiliation{Kavli IPMU (WPI), UTIAS, The University of Tokyo, 5-1-5 Kashiwanoha, Kashiwa, Chiba 277-8583, Japan}
\affiliation{Center for Data-Driven Discovery, Kavli IPMU (WPI), UTIAS, The University of Tokyo, Kashiwa, Chiba 277-8583, Japan}

\date{\today}

\begin{abstract}
Recent observations by the James Webb Space Telescope have revealed massive galaxies at very high redshift ($z\simeq 7-15$). The question of whether the existence of such galaxies is expected in the corresponding JWST surveys has received a lot of attention, though the answer straddles areas of cosmology and complex astrophysical details of high-redshift galaxy formation. The growth rate of density fluctuations determines the  amplitude of overdensities that collapse to form galaxies. Late-time modifications of growth, combined with measurements at both $z\sim 1$ from large-scale structure and $z\sim 1000$ from the cosmic microwave background, affect the predictions for the abundance of first galaxies in the universe. In this paper, we point out that the late-time growth rate of structure affects the statistical significance of high-redshift, high-mass objects very weakly. Consequently, if the existence and abundance of these objects are confirmed to be unexpected, the variations in the late-time growth history are unlikely to explain these anomalies. 
\end{abstract}

\maketitle

\section{Introduction}

The observation of spectroscopically-confirmed, high-redshift, high-mass galaxies by the James Webb Space Telescope \cite{Labbe:2022ahb,Harikane:2022rqt,Naidu2022ApJl,Atek,Casey,Donnan_2022} has caused excitement in astrophysics. Does the standard cosmological model allow for such objects to be created mere hundreds of millions of years after the Big Bang? The answer to this question surely depends on the knotty details of high-redshift galaxy formation (\cite{Blumenthal:1984bp,2023arXiv230411911P,2024arXiv241204983T}), stellar formation \cite{2023OJAp....6E..47M,2024arXiv240618352H,2024Natur.635..311X,2025ApJ...978...89H}, dust physics \cite{2023MNRAS.525.3254S,2025MNRAS.536.1018L} and their interplays \cite{Mirocha2023MNRAS.519..843M,ferrara2023MNRAS.522.3986F,Mason2023MNRAS.521..497M}, all of whose details are not yet well understood. Nevertheless, there have been numerous attempts to quantify the probability of these high-redshift, high-mass events in the standard cosmological model (\cite{Heather,Carnall}) and claims that these objects are at some level of tension with the standard cosmological model 
\cite{Lovell:2022bhx,2023NatAs...7..731B,2024MNRAS.533.3923S}.

One rather obvious yet relatively unexplored question is how the abundance of high-redshift objects observed by JWST is affected by the growth of cosmic structure. Clearly, a higher growth rate (starting from some fixed amplitude of primordial fluctuations) would imply more $z\sim 10$ objects of high mass. However the amplitude of structure growth is constrained not only by the CMB at $z\simeq 1000$, but also by measurements that constrain the amplitude of the matter power spectrum at $z\simeq 0$-$2$. Therefore, the space of possibilities for enhanced growth at $z\simeq 10$ is limited, barring very unusual scenarios where growth would break from the expected scaling with time in a matter-dominated model to be temporarily enhanced around $z\simeq 10$, but then somehow slow back down to its expected scaling with time by $z\simeq 2$.

In this paper we quantitatively address the question of how the growth of cosmic structure affects the predicted abundance of JWST galaxies. We do not attempt to perform a comprehensive parameter search, nor are we particularly concerned about high-accuracy quantification of the rareness of high-redshift galaxies. Rather, we study the change in the standard statistical measures appropriate for the abundance of rare objects when the growth of structure is smoothly varied. We illustrate our results on a few representative examples, and argue that our results strongly indicate that structure growth does not appreciably affect the statistical significance of high-redshift, high-mass galaxies.

\section{Methodology and Data}

The linear growth of structure is described by the function $D(a)$, and further by the growth rate $f(a)\equiv d\ln D/d\ln a$, where $a$ is the scale factor. We make use of the fitting function 
\begin{equation}
    f(a)\simeq \Om^{\gamma}(a),
\end{equation}
where $\gamma$ is the growth index \cite{Linder:2005in}. {For a broad range of possible expansion histories within general relativity, one finds} $\gamma{\simeq}0.55$ with a very weak dependence on the dark energy model \cite{Linder:2007hg}; the linear growth factor is consequently approximated by $D(\gamma, a)=\exp[-\int^1_a d\ln a'\,\Om^{\gamma}(a')]$. We will use the growth index as a tunable parameter that controls the amount of growth at late times. Note the appropriate limits: as $\gamma\rightarrow 0$, $f(a)\rightarrow 1$ and one recovers the growth rate in an Einstein-de Sitter ($\Om=1$) universe; as $\gamma\rightarrow \infty$, $f(a)\rightarrow 0$ and the growth rate is entirely suppressed at late times, when the matter density $\Om(a)$ is below unity. 

Note that late-time changes to growth enabled by the growth-index parameterization do in principle affect the abundance of galaxies even in the epoch before dark energy becomes significant and the effects of $\gamma$ "turn on" (so at $z\gg 1$). This is because low-redshift data that constrain the amplitude of mass fluctuations effectively normalize the growth at low redshift\footnote{This low-redshift normalization is partial, as it is combined with the $z\simeq 1000$ normalization that comes from the CMB in cases when the CMB data are used.}, and thus late-time changes to the growth rate automatically impact the overall growth amplitude, and thus the abundance of objects, at arbitrarily high redshift. In more detail, normalizing the amplitude of mass fluctuations at the present time,
\begin{equation}
    \sigma_8(\gamma, a)=\sigma_8 D(\gamma, a),
\end{equation}
constraints on the present-day amplitude of mass fluctuations $\sigma_8$, along with the growth model parametrized by the growth index $\gamma$, together affect the amplitude of mass fluctuations and hence the abundance of galaxies at all times.

To quantify the probability of high-mass high-redshift galaxies, we use the extreme value statistics \cite{Gumbel+1958} which has been applied in this context previously \cite{Lovell:2022bhx}. The starting point is the stellar probability distribution function (PDF), $\Phi(M^*_{\rm max})$, for the most massive galaxy, which is the product of the PDF of the most massive halo $\Phi(M^{\rm DM}_{\rm max})$, the baryon fraction $f_b$, and the stellar fractions $f_*$
\begin{equation}
    \Phi(M^*_{\rm max})=\Phi(M^{\rm DM}_{\rm max})f_bf_*.
\end{equation}
Here we assume $f_*$ has a truncated lognormal distribution $f_*=\ln N(\mu,\sigma^2)$,
where $\mu=e^{-2}$ and $\sigma=1$, all chosen in \cite{Lovell:2022bhx} so as to approximately match findings from a combination of halo models and observations. 
In most of our tests, we leave $\gamma$ free while fixing all the $\Lambda$CDM cosmological parameters to their Planck \cite{Planck:2018vyg} values, specifically the scaled Hubble constant $h=0.673$, physical matter and baryon densities $\Omega_{m}h^2=0.143$ and $\Omega_{b}h^2=0.022$, present-day amplitude of mass fluctuations $\sigma_8=0.811$, and the scalar spectral index $n_s=0.965$; these values also fix the baryon fraction to $f_b\equiv \Omega_b h^2 / (\Omega_c h^2 + \Omega_b h^2) = 0.16$. Near the end of our Results section, we show a test on models selected from a Markov Chain in which all of the cosmological parameters are allowed to vary.

{
Firstly, consider a survey covering a sky fraction $\fsky$ over the redshift interval $[z_{\min}, z_{\max}]$.  The total expected number of halos in this volume is
\begin{equation}
    n_{\text{tot}}=\fsky \left [\int^{z_{\max}}_{z_{\min}}\int^{\infty}_{0}dzdM\frac{dV}{dz}\frac{dn(M,z)}{dM} \right ]
\end{equation}
To derive the extreme‐value statistics, we bin the redshift range in intervals of width $\Delta z = 0.2$ and compute the mass‐distribution PDF within each bin.  The normalized PDF of halo masses is then
\begin{equation}
    g(m)=\frac{f_{\text{sky}}}{n_{\text{tot}}}\left [\int^{z_{\max}}_{z_{\min}}dz\frac{dV}{dz}\frac{dn(m,z)}{dm}\right ],
    \label{eq:g}
\end{equation}
which gives the probability density for finding a halo of mass $M$ in the specified survey volume.}
By integrating over $M$, the cumulative distribution function (CDF) is
\begin{equation}
    G(m)=\frac{f_{\text{sky}}}{n_{\text{tot}}}\left [\int^{z_{\max}}_{z_{\min}}\int^m_{{0}}dMdz\frac{dV}{dz}\frac{dn(M,z)}{dM}\right ].
    \label{eq:G}
\end{equation}
We then consider the distribution of halo masses within a given volume as a sequence of independent and identically distributed random variables drawn from the distribution described above, $\{M_1,\dots,M_{n_{tot}}\}$ . The probability that all of these variables are less than or equal to a value $m$ is given by the product of the CDF of the halo mass distribution
\begin{equation}
    \Phi(M_{\rm max}^{\rm DM}\leq m)=G(m)^{n_{\rm tot}}
    \label{CDF_Phi}
\end{equation}
where $M_{\rm max}^{\rm DM}$ is the largest value of the sequence. Taking the derivative of \eqref{CDF_Phi} yields the probability of the most massive halo to have a mass of $m$:
\begin{equation}
    \Phi(M^{\rm DM}_{\rm max}=m)=n_{\rm tot}g(m)G(m)^{n_{\rm tot}-1}.
\end{equation}
Note that we do not vary some of the imprecisely known inputs to this formalism (for example, the stellar fraction $f_*$) since our goal is not to quantify the prediction for the highest-mass object, but rather to study its dependence on the growth of structure.

An important ingredient required in the prescription in Eq.~(\ref{eq:g}-\ref{eq:G}) is the mass function $dn/dM$. Most efforts of calibrating the mass function have been performed at lower redshift (\cite{Sheth:2001dp,Jenkins:2000bv,Tinker:2008ff}), but there does exist a body of literature that has specifically targeted the $z\sim 10$ range \cite{Reed:2006rw}. Here, we adopt the \citet{Warren:2005ey} mass function, in which the halo multiplicity function is written as
\begin{equation}
f(\sigma)=0.7234(\sigma^{-1.625}+0.2538)\exp[-\frac{1.1982}{\sigma^2}],
\end{equation} {where $\sigma$ is the root-mean-square(rms) amplitude of the matter overdensity field smoothed on the spatial scale that encloses mass $M$. \citet{Lukic:2007fc} have shown that this mass function remains valid even at high redshift. We have also explored the result from different halo mass functions in the following section.}

Finally, we need to correct the observed masses for Eddington Bias (\cite{Eddington}) --- the fact that when objects with uncertain mass are selected from a steeply falling mass function, it is more likely that a low-mass halo is "scattered" to a higher mass than the other way around. The Eddington bias-corrected mass is (e.g.\ \cite{Stanek:2006tu,Mortonson:2010mj})
\begin{equation}
    \ln M_{\text{edd}}=\ln M_{\text{obs}}+\frac{1}{2}\epsilon\sigma^2_{\ln M}
    \label{eq:Eddington}
\end{equation}
where $M_{\text{obs}}$ is the mass reported by the observations, $\epsilon$ is the local slope of the underlying halo mass function, and $\sigma_{\ln M}$ is the uncertainty in the halo/stellar mass estimate. The correction has the familiar property of being proportional to both the slope of the mass function and the variance in the measurement of mass of the object. 

Finally, we integrate over the PDF for the most massive object in order to obtain the prediction for its mass. In more detail, we integrate $\Phi(M^*_{\rm max})$ over $M$ in order to determine the predicted 1, 2, and 3-sigma probability ranges for mass of the most extreme object in the survey. Given measurements of the masses and redshifts of a sample of objects, we can in turn obtain the statistical significance of the reported mass of the most massive object at a given redshift.

As far as the data are concerned, we do not attempt to be comprehensive, but we have checked that the conclusions are unchanged when different datasets are considered. We select the most extreme object in a given survey, by which we mean the object which, in the $z-\log(M)$ plane, most deviates from expectations from extreme value theory (we do not show those expectations, but they are straightforwardly computed using the formalism we lay out above). To be specific, we select the most deviant and spectroscopically confirmed such object in the \citet{2024Natur.635..311X} sample, which is the galaxy  \texttt{S1} at $z=5.58$ with the stellar mass of $\log(M^*/\Msun)=11.37_{-0.13}^{+0.11}$.  

\begin{figure}[!t]
    \centering
\includegraphics[width=\linewidth]{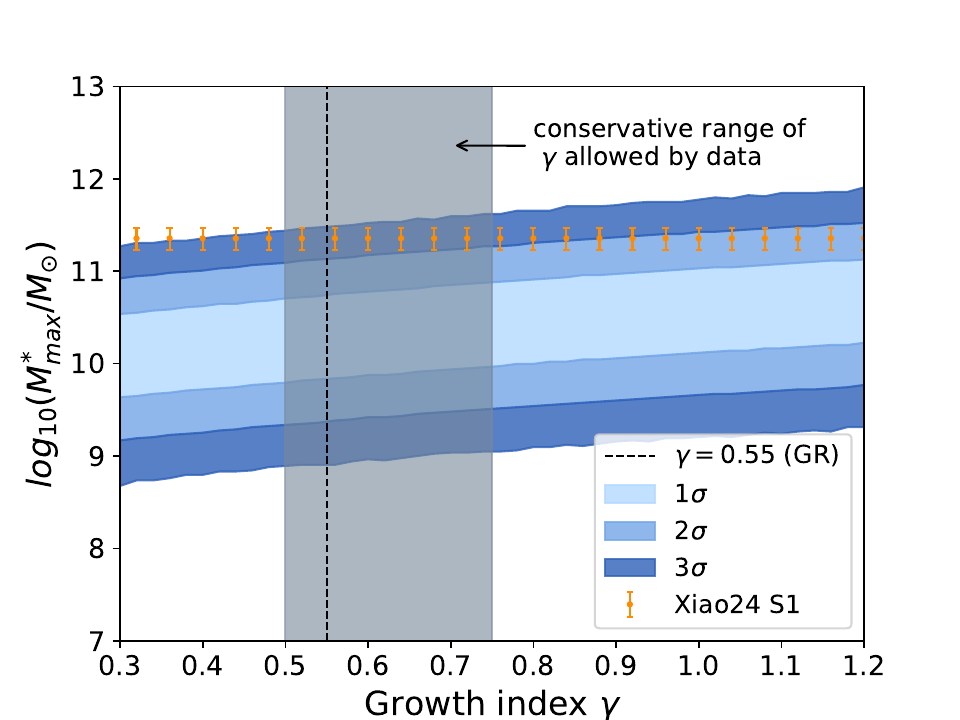}
\caption{Logarithm of the most massive object’s mass ($y$-axis) expected in the survey, based on the specifications of the \citet{2024Natur.635..311X} sample, as a function of the growth index $\gamma$ ($x$-axis). The horizontal colored bands show the 68.3\%, 95.4\%, and 99.7\% credible intervals for the mass of the highest-mass object in that sample, as a function of $\gamma$.  The orange error bar (independent of the theory parameter on the x-axis, and shown multiple times for viewing convenience) shows the actual measurement of the highest-mass object in this sample at $z=5.58$ in this sample, corrected for Eddington bias. The vertical band shows the $\pm$5$\sigma$ range of values of $\gamma$ allowed by present data.
}
\label{fig:logM_vs_gamma_Xiao}
\end{figure}

\section{Results}

Our principal result is shown in Fig.~\ref{fig:logM_vs_gamma_Xiao}. Here we show the predicted mass of the aforementioned object \texttt{S1} at $z=5.58$. To quantify the significance of its mass measured at this redshift, we specialize in the redshift bin $z=5.58\pm 0.10$, and consider the expectation for most massive object in a survey of $124$ square arcminutes. In the Figure, the horizontal bands show the 68.3\%, 95.4\%, and 99.7\% credible intervals for the \textit{expected} mass of the highest-mass object at $z\simeq 5.6$ in the \citet{2024Natur.635..311X} sample, as a function of $\gamma$. The orange error bar shows the \textit{actual} measurement of the highest-mass object in the sample, the aforementioned \texttt{S1}. Note that we have repeated showing this mass measurement at a number of values of $\gamma$ in order to illustrate where the measurement lies relative to expectation (horizontal bands) for any arbitrary cosmological model parameterized by the growth index.

The principal takeaway from Fig.~\ref{fig:logM_vs_gamma_Xiao} is the slow dependence of the expected mass limit as a function of $\gamma$. To help see this, the vertical range in the Figure shows the approximate and \textit{very conservative} range of values of the growth index allowed by the data, corresponding to $\pm 5$$\sigma$ range from current data\footnote{We adopt the cosmological-constraint results from \cite{Nguyen:2023fip}; the 5$\sigma$ range allowed is approximately $0.50<\gamma< 0.75$, which corresponds to the range that we have selected. We hold the other cosmological parameters fixed in this estimate, but allow them to vary further down in this analysis.}. For example, for $\gamma = 0.50$, we find that the galaxy \texttt{S1} is higher than its predicted mass range at the significance of $2.7\sigma$, while for $\gamma=0.75$ (and all other parameters unchanged), this changes to $2.2\sigma$. While this change in significance is not entirely negligible, it is very modest given the big change in the growth rate encoded by varying the growth index between these two values. {We have repeated the analysis with the Reed \cite{Reed:2006rw}, Tinker \cite{Tinker:2008ff} and Press-Schechter \cite{1974ApJ...187..425P} mass functions, given the significance range of [2.55$\sigma$, 2.02$\sigma$], [2.78$\sigma$, 2.27$\sigma$], [3.12$\sigma$, 2.60$\sigma$] respectively as $\gamma$ varies from 0.50 to 0.75, confirming that our conclusions remain robust.}

We have checked that the weak dependence is even more pronounced for the highest-mass objects that are not, at face value, unexpected according to extreme-value statistics computation. For example in the \citet{Casey} sample, after correcting the Eddington Bias, the galaxy \texttt{COS-z13-2} ($M_*=(5.6^{+3.4}_{-2.2})\times10^9M_{\odot}$, $z=13.4^{+0.7}_{-1.2}$) is more massive than expected with significance of $3.1\sigma$ when $\gamma=0.5$, and the value goes down to just $2.3\sigma$  when $\gamma=0.75$. In Labbe\cite{Labbe:2022ahb} sample, the galaxy \texttt{id38094} ($\log(M_*/M_{\odot})=10.89^{+0.09}_{-0.08}$, $z=7.48^{+0.04}_{-0.04}$)  has significance of $3.3\sigma$ at $\gamma=0.50$, which only goes down to $2.8\sigma$ when $\gamma=0.75$. 
{For completeness, we have also checked that the JADES-GS-z14-0 galaxy, with mass $\log(M_*/M_{\odot})=8.6^{+0.7}_{-0.2}$ and redshift $z=14.32^{+0.08}_{-0.20}$ \citep{2024Natur.633..318C}---most distant galaxy yet spectroscopically confirmed---is consistent with $\Lambda$CDM: assuming $\gamma$ = 0.55 its significance is 0.76$\sigma$, and as $\gamma$ varies from 0.50 to 0.75 the significance ranges from 0.93$\sigma$ to 0.37$\sigma$.}

A more representative illustration of the effect of the growth of structure (than a change in one parameter that holds all others fixed) may be obtained by comparing the expectation for the extreme-value statistic in a range of models consistent with current data. To that effect, we take the $\Lambda$CDM Markov chain with a varying gamma from the cosmological analysis in \citet{Nguyen:2023fip}, which assumes the combined data of temperature, polarization, and lensing from Planck \cite{Planck:2018vyg}, the combined galaxy clustering and weak lensing ("$3\times 2$") analysis from the first year of data from Dark Energy Survey \cite{DESY1:3x2pt},  baryon acoustic oscillations (BAO) in the 6dF Galaxy Survey (6dFGS) galaxy \cite{6dF:BAO_H0} and the Sloan Digital Sky Survey (SDSS) \cite{SDSS:DR7,SDSS:DR12,SDSS:DR16}, and redshift-space distortion (RSD) constraints on the growth of structure at local ($z<0.1$) \cite{6dF:growth_sigma8,Huterer:2016uyq,Said:2020,Boruah:2020,Turner:2023} and cosmological distances ($z\geq0.1$) \cite{Blake:2011,Blake:2013,Howlett:2015,Okumura:2016,Pezzotta:2017,SDSS:DR16}. The dataset we adopt corresponds to the third row of Table 1 in \cite{Nguyen:2023fip}. For each model in the chain, we compute the range of expectations for the mass (as we did in Fig.~\ref{fig:logM_vs_gamma_Xiao} for varying the growth index alone), and evaluate the significance of the mass of galaxy \texttt{S1}. We then calculate the range of these significances from the chain. To be conservative, we quote the 3$\sigma$ range
\begin{equation}
    \mbox{significance}\in [2.34\sigma, 2.73\sigma]\quad ({\rm at}\,\, 99.7\%)
\end{equation}
This more realistic example shows the very mild variation in the significance of the existence of a high-redshift, high-mass galaxy when we allow the variation in the growth of cosmic structure even beyond those allowed in the $\Lambda$CDM model.

\section{Conclusions}
Recent JWST observations have uncovered unexpectedly massive galaxies at high redshift, but it remains unclear whether their high masses are due to unexpected features in the cosmological model, or more-complex-than-expected astrophysics of galaxy formation at high redshift. 

Variations in the late-time growth rate, combined with low-redshift measurements that are sensitive to the amplitude of mass fluctuations, imply modified expectations for the abundance of objects at arbitrarily high redshift. Thus far, the magnitude of the impact of late-time growth variations on the predicted abundance of high-redshift objects has not been quantified. This is where the present work comes in: we examine, for the first time, the impact of late-time growth history on the predicted masses of the most massive observed galaxies at a given redshift.  

By applying extreme-value statistics to a few selected observational samples, we demonstrated that even significant changes in the late-time growth of structure, described by the "growth index" parametrization, lead to only modest changes in the statistical significance of the most massive observed galaxies. For example, a very large ({5$\sigma$}, based on current data) shift in the growth index leads to the change in the reported significance of the most extreme galaxy \texttt{S1}  \cite{2024Natur.635..311X} from 2.73$\sigma$ to just 2.23$\sigma$. We supplied several more examples, including a case where all cosmological parameters were varied within observational bounds, to illustrate the weak dependence of the theoretical expectations for the rareness of objects as a function of cosmic growth. 

One possible caveat to our findings is that we only consider smooth changes in the growth of structure, such as those described by the growth index $\gamma$. Though it seems unlikely, it is possible that a more rapid onset of the growth of structure prior to $z\simeq 10$ could significantly boost the predicted abundance of galaxies, and thus more strongly change the extreme-value statistics adopted in this paper. We are not aware of any realistic physical models that would enable such a sudden onset of growth, but it is important to keep this possibility in mind should the abundance of high-redshift, high-mass objects raise to the level of a cosmological tension. Yet another caveat is that we only consider structure growth from (almost) scale-invariant, Gaussian primordial fluctuations. Alternatives, including a blue-tilted \cite{Hirano:2015wla,Hirano:2023auh,Parashari:2023cui} or (strongly) non-Gaussian-and-scale-dependent \cite{Biagetti:2022ode} primordial power spectrum, can allow for early onset of nonlinear structure growth at scales relevant for the formation of these high-redshift, high-mass galaxies found by JWST. Such scenarios, however, remain undetected with current observations.

\section*{Acknowledgments}
We thank Christopher Lovell and Peter Behroozi for helpful discussions. This work has been supported by the Department of Energy under contract DE‐SC0019193 and Leinweber Center for Theoretical Physics at the University of Michigan. 

\bibliographystyle{apsrev4-2}
\bibliography{references}

\begin{thebibliography}{58}%
\makeatletter
\providecommand \@ifxundefined [1]{%
 \@ifx{#1\undefined}
}%
\providecommand \@ifnum [1]{%
 \ifnum #1\expandafter \@firstoftwo
 \else \expandafter \@secondoftwo
 \fi
}%
\providecommand \@ifx [1]{%
 \ifx #1\expandafter \@firstoftwo
 \else \expandafter \@secondoftwo
 \fi
}%
\providecommand \natexlab [1]{#1}%
\providecommand \enquote  [1]{``#1''}%
\providecommand \bibnamefont  [1]{#1}%
\providecommand \bibfnamefont [1]{#1}%
\providecommand \citenamefont [1]{#1}%
\providecommand \href@noop [0]{\@secondoftwo}%
\providecommand \href [0]{\begingroup \@sanitize@url \@href}%
\providecommand \@href[1]{\@@startlink{#1}\@@href}%
\providecommand \@@href[1]{\endgroup#1\@@endlink}%
\providecommand \@sanitize@url [0]{\catcode `\\12\catcode `\$12\catcode `\&12\catcode `\#12\catcode `\^12\catcode `\_12\catcode `\%12\relax}%
\providecommand \@@startlink[1]{}%
\providecommand \@@endlink[0]{}%
\providecommand \url  [0]{\begingroup\@sanitize@url \@url }%
\providecommand \@url [1]{\endgroup\@href {#1}{\urlprefix }}%
\providecommand \urlprefix  [0]{URL }%
\providecommand \Eprint [0]{\href }%
\providecommand \doibase [0]{https://doi.org/}%
\providecommand \selectlanguage [0]{\@gobble}%
\providecommand \bibinfo  [0]{\@secondoftwo}%
\providecommand \bibfield  [0]{\@secondoftwo}%
\providecommand \translation [1]{[#1]}%
\providecommand \BibitemOpen [0]{}%
\providecommand \bibitemStop [0]{}%
\providecommand \bibitemNoStop [0]{.\EOS\space}%
\providecommand \EOS [0]{\spacefactor3000\relax}%
\providecommand \BibitemShut  [1]{\csname bibitem#1\endcsname}%
\let\auto@bib@innerbib\@empty
\bibitem [{\citenamefont {Labbe}\ \emph {et~al.}(2023)\citenamefont {Labbe} \emph {et~al.}}]{Labbe:2022ahb}%
  \BibitemOpen
  \bibfield  {author} {\bibinfo {author} {\bibfnamefont {I.}~\bibnamefont {Labbe}} \emph {et~al.},\ }\href {https://doi.org/10.1038/s41586-023-05786-2} {\bibfield  {journal} {\bibinfo  {journal} {Nature}\ }\textbf {\bibinfo {volume} {616}},\ \bibinfo {pages} {266} (\bibinfo {year} {2023})},\ \Eprint {https://arxiv.org/abs/2207.12446} {arXiv:2207.12446 [astro-ph.GA]} \BibitemShut {NoStop}%
\bibitem [{\citenamefont {Harikane}\ \emph {et~al.}(2023)\citenamefont {Harikane}, \citenamefont {Ouchi}, \citenamefont {Oguri}, \citenamefont {Ono}, \citenamefont {Nakajima}, \citenamefont {Isobe}, \citenamefont {Umeda}, \citenamefont {Mawatari},\ and\ \citenamefont {Zhang}}]{Harikane:2022rqt}%
  \BibitemOpen
  \bibfield  {author} {\bibinfo {author} {\bibfnamefont {Y.}~\bibnamefont {Harikane}}, \bibinfo {author} {\bibfnamefont {M.}~\bibnamefont {Ouchi}}, \bibinfo {author} {\bibfnamefont {M.}~\bibnamefont {Oguri}}, \bibinfo {author} {\bibfnamefont {Y.}~\bibnamefont {Ono}}, \bibinfo {author} {\bibfnamefont {K.}~\bibnamefont {Nakajima}}, \bibinfo {author} {\bibfnamefont {Y.}~\bibnamefont {Isobe}}, \bibinfo {author} {\bibfnamefont {H.}~\bibnamefont {Umeda}}, \bibinfo {author} {\bibfnamefont {K.}~\bibnamefont {Mawatari}},\ and\ \bibinfo {author} {\bibfnamefont {Y.}~\bibnamefont {Zhang}},\ }\bibfield  {journal} {\bibinfo  {journal} {Astrophys. J. Suppl.}\ }\textbf {\bibinfo {volume} {265}},\ \href {https://doi.org/10.3847/1538-4365/acaaa9} {10.3847/1538-4365/acaaa9} (\bibinfo {year} {2023}),\ \Eprint {https://arxiv.org/abs/2208.01612} {arXiv:2208.01612 [astro-ph.GA]} \BibitemShut {NoStop}%
\bibitem [{\citenamefont {{Naidu}}\ \emph {et~al.}(2022)\citenamefont {{Naidu}}, \citenamefont {{Oesch}}, \citenamefont {{van Dokkum}}, \citenamefont {{Nelson}}, \citenamefont {{Suess}}, \citenamefont {{Brammer}}, \citenamefont {{Whitaker}}, \citenamefont {{Illingworth}}, \citenamefont {{Bouwens}}, \citenamefont {{Tacchella}}, \citenamefont {{Matthee}}, \citenamefont {{Allen}}, \citenamefont {{Bezanson}}, \citenamefont {{Conroy}}, \citenamefont {{Labbe}}, \citenamefont {{Leja}}, \citenamefont {{Leonova}}, \citenamefont {{Magee}}, \citenamefont {{Price}}, \citenamefont {{Setton}}, \citenamefont {{Strait}}, \citenamefont {{Stefanon}}, \citenamefont {{Toft}}, \citenamefont {{Weaver}},\ and\ \citenamefont {{Weibel}}}]{Naidu2022ApJl}%
  \BibitemOpen
  \bibfield  {author} {\bibinfo {author} {\bibfnamefont {R.~P.}\ \bibnamefont {{Naidu}}}, \bibinfo {author} {\bibfnamefont {P.~A.}\ \bibnamefont {{Oesch}}}, \bibinfo {author} {\bibfnamefont {P.}~\bibnamefont {{van Dokkum}}}, \bibinfo {author} {\bibfnamefont {E.~J.}\ \bibnamefont {{Nelson}}}, \bibinfo {author} {\bibfnamefont {K.~A.}\ \bibnamefont {{Suess}}}, \bibinfo {author} {\bibfnamefont {G.}~\bibnamefont {{Brammer}}}, \bibinfo {author} {\bibfnamefont {K.~E.}\ \bibnamefont {{Whitaker}}}, \bibinfo {author} {\bibfnamefont {G.}~\bibnamefont {{Illingworth}}}, \bibinfo {author} {\bibfnamefont {R.}~\bibnamefont {{Bouwens}}}, \bibinfo {author} {\bibfnamefont {S.}~\bibnamefont {{Tacchella}}}, \bibinfo {author} {\bibfnamefont {J.}~\bibnamefont {{Matthee}}}, \bibinfo {author} {\bibfnamefont {N.}~\bibnamefont {{Allen}}}, \bibinfo {author} {\bibfnamefont {R.}~\bibnamefont {{Bezanson}}}, \bibinfo {author} {\bibfnamefont {C.}~\bibnamefont {{Conroy}}}, \bibinfo {author} {\bibfnamefont {I.}~\bibnamefont {{Labbe}}},
  \bibinfo {author} {\bibfnamefont {J.}~\bibnamefont {{Leja}}}, \bibinfo {author} {\bibfnamefont {E.}~\bibnamefont {{Leonova}}}, \bibinfo {author} {\bibfnamefont {D.}~\bibnamefont {{Magee}}}, \bibinfo {author} {\bibfnamefont {S.~H.}\ \bibnamefont {{Price}}}, \bibinfo {author} {\bibfnamefont {D.~J.}\ \bibnamefont {{Setton}}}, \bibinfo {author} {\bibfnamefont {V.}~\bibnamefont {{Strait}}}, \bibinfo {author} {\bibfnamefont {M.}~\bibnamefont {{Stefanon}}}, \bibinfo {author} {\bibfnamefont {S.}~\bibnamefont {{Toft}}}, \bibinfo {author} {\bibfnamefont {J.~R.}\ \bibnamefont {{Weaver}}},\ and\ \bibinfo {author} {\bibfnamefont {A.}~\bibnamefont {{Weibel}}},\ }\href {https://doi.org/10.3847/2041-8213/ac9b22} {\bibfield  {journal} {\bibinfo  {journal} {\apjl}\ }\textbf {\bibinfo {volume} {940}},\ \bibinfo {eid} {L14} (\bibinfo {year} {2022})},\ \Eprint {https://arxiv.org/abs/2207.09434} {arXiv:2207.09434 [astro-ph.GA]} \BibitemShut {NoStop}%
\bibitem [{\citenamefont {{Atek}}\ \emph {et~al.}(2023)\citenamefont {{Atek}}, \citenamefont {{Shuntov}}, \citenamefont {{Furtak}}, \citenamefont {{Richard}}, \citenamefont {{Kneib}}, \citenamefont {{Mahler}}, \citenamefont {{Zitrin}}, \citenamefont {{McCracken}}, \citenamefont {{Charlot}}, \citenamefont {{Chevallard}},\ and\ \citenamefont {{Chemerynska}}}]{Atek}%
  \BibitemOpen
  \bibfield  {author} {\bibinfo {author} {\bibfnamefont {H.}~\bibnamefont {{Atek}}}, \bibinfo {author} {\bibfnamefont {M.}~\bibnamefont {{Shuntov}}}, \bibinfo {author} {\bibfnamefont {L.~J.}\ \bibnamefont {{Furtak}}}, \bibinfo {author} {\bibfnamefont {J.}~\bibnamefont {{Richard}}}, \bibinfo {author} {\bibfnamefont {J.-P.}\ \bibnamefont {{Kneib}}}, \bibinfo {author} {\bibfnamefont {G.}~\bibnamefont {{Mahler}}}, \bibinfo {author} {\bibfnamefont {A.}~\bibnamefont {{Zitrin}}}, \bibinfo {author} {\bibfnamefont {H.~J.}\ \bibnamefont {{McCracken}}}, \bibinfo {author} {\bibfnamefont {S.}~\bibnamefont {{Charlot}}}, \bibinfo {author} {\bibfnamefont {J.}~\bibnamefont {{Chevallard}}},\ and\ \bibinfo {author} {\bibfnamefont {I.}~\bibnamefont {{Chemerynska}}},\ }\href {https://doi.org/10.1093/mnras/stac3144} {\bibfield  {journal} {\bibinfo  {journal} {Mon. Not. Roy. Astron. Soc.}\ }\textbf {\bibinfo {volume} {519}},\ \bibinfo {pages} {1201} (\bibinfo {year} {2023})},\ \Eprint {https://arxiv.org/abs/2207.12338}
  {arXiv:2207.12338 [astro-ph.GA]} \BibitemShut {NoStop}%
\bibitem [{\citenamefont {{Casey}}\ \emph {et~al.}(2024)\citenamefont {{Casey}}, \citenamefont {{Akins}}, \citenamefont {{Shuntov}}, \citenamefont {{Ilbert}}, \citenamefont {{Paquereau}}, \citenamefont {{Franco}}, \citenamefont {{Hayward}}, \citenamefont {{Finkelstein}}, \citenamefont {{Boylan-Kolchin}}, \citenamefont {{Robertson}}, \citenamefont {{Allen}}, \citenamefont {{Brinch}}, \citenamefont {{Cooper}}, \citenamefont {{Ding}}, \citenamefont {{Drakos}}, \citenamefont {{Faisst}}, \citenamefont {{Fujimoto}}, \citenamefont {{Gillman}}, \citenamefont {{Harish}}, \citenamefont {{Hirschmann}}, \citenamefont {{Jin}}, \citenamefont {{Kartaltepe}}, \citenamefont {{Koekemoer}}, \citenamefont {{Kokorev}}, \citenamefont {{Liu}}, \citenamefont {{Long}}, \citenamefont {{Magdis}}, \citenamefont {{Maraston}}, \citenamefont {{Martin}}, \citenamefont {{McCracken}}, \citenamefont {{McKinney}}, \citenamefont {{Mobasher}}, \citenamefont {{Rhodes}}, \citenamefont {{Rich}}, \citenamefont {{Sanders}}, \citenamefont
  {{Silverman}}, \citenamefont {{Toft}}, \citenamefont {{Vijayan}}, \citenamefont {{Weaver}}, \citenamefont {{Wilkins}}, \citenamefont {{Yang}},\ and\ \citenamefont {{Zavala}}}]{Casey}%
  \BibitemOpen
  \bibfield  {author} {\bibinfo {author} {\bibfnamefont {C.~M.}\ \bibnamefont {{Casey}}}, \bibinfo {author} {\bibfnamefont {H.~B.}\ \bibnamefont {{Akins}}}, \bibinfo {author} {\bibfnamefont {M.}~\bibnamefont {{Shuntov}}}, \bibinfo {author} {\bibfnamefont {O.}~\bibnamefont {{Ilbert}}}, \bibinfo {author} {\bibfnamefont {L.}~\bibnamefont {{Paquereau}}}, \bibinfo {author} {\bibfnamefont {M.}~\bibnamefont {{Franco}}}, \bibinfo {author} {\bibfnamefont {C.~C.}\ \bibnamefont {{Hayward}}}, \bibinfo {author} {\bibfnamefont {S.~L.}\ \bibnamefont {{Finkelstein}}}, \bibinfo {author} {\bibfnamefont {M.}~\bibnamefont {{Boylan-Kolchin}}}, \bibinfo {author} {\bibfnamefont {B.~E.}\ \bibnamefont {{Robertson}}}, \bibinfo {author} {\bibfnamefont {N.}~\bibnamefont {{Allen}}}, \bibinfo {author} {\bibfnamefont {M.}~\bibnamefont {{Brinch}}}, \bibinfo {author} {\bibfnamefont {O.~R.}\ \bibnamefont {{Cooper}}}, \bibinfo {author} {\bibfnamefont {X.}~\bibnamefont {{Ding}}}, \bibinfo {author} {\bibfnamefont {N.~E.}\ \bibnamefont
  {{Drakos}}}, \bibinfo {author} {\bibfnamefont {A.~L.}\ \bibnamefont {{Faisst}}}, \bibinfo {author} {\bibfnamefont {S.}~\bibnamefont {{Fujimoto}}}, \bibinfo {author} {\bibfnamefont {S.}~\bibnamefont {{Gillman}}}, \bibinfo {author} {\bibfnamefont {S.}~\bibnamefont {{Harish}}}, \bibinfo {author} {\bibfnamefont {M.}~\bibnamefont {{Hirschmann}}}, \bibinfo {author} {\bibfnamefont {S.}~\bibnamefont {{Jin}}}, \bibinfo {author} {\bibfnamefont {J.~S.}\ \bibnamefont {{Kartaltepe}}}, \bibinfo {author} {\bibfnamefont {A.~M.}\ \bibnamefont {{Koekemoer}}}, \bibinfo {author} {\bibfnamefont {V.}~\bibnamefont {{Kokorev}}}, \bibinfo {author} {\bibfnamefont {D.}~\bibnamefont {{Liu}}}, \bibinfo {author} {\bibfnamefont {A.~S.}\ \bibnamefont {{Long}}}, \bibinfo {author} {\bibfnamefont {G.}~\bibnamefont {{Magdis}}}, \bibinfo {author} {\bibfnamefont {C.}~\bibnamefont {{Maraston}}}, \bibinfo {author} {\bibfnamefont {C.~L.}\ \bibnamefont {{Martin}}}, \bibinfo {author} {\bibfnamefont {H.~J.}\ \bibnamefont {{McCracken}}}, \bibinfo
  {author} {\bibfnamefont {J.}~\bibnamefont {{McKinney}}}, \bibinfo {author} {\bibfnamefont {B.}~\bibnamefont {{Mobasher}}}, \bibinfo {author} {\bibfnamefont {J.}~\bibnamefont {{Rhodes}}}, \bibinfo {author} {\bibfnamefont {R.~M.}\ \bibnamefont {{Rich}}}, \bibinfo {author} {\bibfnamefont {D.~B.}\ \bibnamefont {{Sanders}}}, \bibinfo {author} {\bibfnamefont {J.~D.}\ \bibnamefont {{Silverman}}}, \bibinfo {author} {\bibfnamefont {S.}~\bibnamefont {{Toft}}}, \bibinfo {author} {\bibfnamefont {A.~P.}\ \bibnamefont {{Vijayan}}}, \bibinfo {author} {\bibfnamefont {J.~R.}\ \bibnamefont {{Weaver}}}, \bibinfo {author} {\bibfnamefont {S.~M.}\ \bibnamefont {{Wilkins}}}, \bibinfo {author} {\bibfnamefont {L.}~\bibnamefont {{Yang}}},\ and\ \bibinfo {author} {\bibfnamefont {J.~A.}\ \bibnamefont {{Zavala}}},\ }\href {https://doi.org/10.3847/1538-4357/ad2075} {\bibfield  {journal} {\bibinfo  {journal} {Astrophys. J.}\ }\textbf {\bibinfo {volume} {965}},\ \bibinfo {eid} {98} (\bibinfo {year} {2024})},\ \Eprint
  {https://arxiv.org/abs/2308.10932} {arXiv:2308.10932 [astro-ph.GA]} \BibitemShut {NoStop}%
\bibitem [{\citenamefont {Donnan}\ \emph {et~al.}(2022)\citenamefont {Donnan}, \citenamefont {McLeod}, \citenamefont {Dunlop}, \citenamefont {McLure}, \citenamefont {Carnall}, \citenamefont {Begley}, \citenamefont {Cullen}, \citenamefont {Hamadouche}, \citenamefont {Bowler}, \citenamefont {Magee}, \citenamefont {McCracken}, \citenamefont {Milvang-Jensen}, \citenamefont {Moneti},\ and\ \citenamefont {Targett}}]{Donnan_2022}%
  \BibitemOpen
  \bibfield  {author} {\bibinfo {author} {\bibfnamefont {C.~T.}\ \bibnamefont {Donnan}}, \bibinfo {author} {\bibfnamefont {D.~J.}\ \bibnamefont {McLeod}}, \bibinfo {author} {\bibfnamefont {J.~S.}\ \bibnamefont {Dunlop}}, \bibinfo {author} {\bibfnamefont {R.~J.}\ \bibnamefont {McLure}}, \bibinfo {author} {\bibfnamefont {A.~C.}\ \bibnamefont {Carnall}}, \bibinfo {author} {\bibfnamefont {R.}~\bibnamefont {Begley}}, \bibinfo {author} {\bibfnamefont {F.}~\bibnamefont {Cullen}}, \bibinfo {author} {\bibfnamefont {M.~L.}\ \bibnamefont {Hamadouche}}, \bibinfo {author} {\bibfnamefont {R.~A.~A.}\ \bibnamefont {Bowler}}, \bibinfo {author} {\bibfnamefont {D.}~\bibnamefont {Magee}}, \bibinfo {author} {\bibfnamefont {H.~J.}\ \bibnamefont {McCracken}}, \bibinfo {author} {\bibfnamefont {B.}~\bibnamefont {Milvang-Jensen}}, \bibinfo {author} {\bibfnamefont {A.}~\bibnamefont {Moneti}},\ and\ \bibinfo {author} {\bibfnamefont {T.}~\bibnamefont {Targett}},\ }\href {https://doi.org/10.1093/mnras/stac3472} {\bibfield  {journal}
  {\bibinfo  {journal} {Mon. Not. Roy. Astron. Soc.}\ }\textbf {\bibinfo {volume} {518}},\ \bibinfo {pages} {6011–6040} (\bibinfo {year} {2022})}\BibitemShut {NoStop}%
\bibitem [{\citenamefont {Blumenthal}\ \emph {et~al.}(1984)\citenamefont {Blumenthal}, \citenamefont {Faber}, \citenamefont {Primack},\ and\ \citenamefont {Rees}}]{Blumenthal:1984bp}%
  \BibitemOpen
  \bibfield  {author} {\bibinfo {author} {\bibfnamefont {G.~R.}\ \bibnamefont {Blumenthal}}, \bibinfo {author} {\bibfnamefont {S.~M.}\ \bibnamefont {Faber}}, \bibinfo {author} {\bibfnamefont {J.~R.}\ \bibnamefont {Primack}},\ and\ \bibinfo {author} {\bibfnamefont {M.~J.}\ \bibnamefont {Rees}},\ }\href {https://doi.org/10.1038/311517a0} {\bibfield  {journal} {\bibinfo  {journal} {Nature}\ }\textbf {\bibinfo {volume} {311}},\ \bibinfo {pages} {517} (\bibinfo {year} {1984})}\BibitemShut {NoStop}%
\bibitem [{\citenamefont {{Prada}}\ \emph {et~al.}(2023)\citenamefont {{Prada}}, \citenamefont {{Behroozi}}, \citenamefont {{Ishiyama}}, \citenamefont {{Klypin}},\ and\ \citenamefont {{P{\'e}rez}}}]{2023arXiv230411911P}%
  \BibitemOpen
  \bibfield  {author} {\bibinfo {author} {\bibfnamefont {F.}~\bibnamefont {{Prada}}}, \bibinfo {author} {\bibfnamefont {P.}~\bibnamefont {{Behroozi}}}, \bibinfo {author} {\bibfnamefont {T.}~\bibnamefont {{Ishiyama}}}, \bibinfo {author} {\bibfnamefont {A.}~\bibnamefont {{Klypin}}},\ and\ \bibinfo {author} {\bibfnamefont {E.}~\bibnamefont {{P{\'e}rez}}},\ }\href {https://doi.org/10.48550/arXiv.2304.11911} {\bibfield  {journal} {\bibinfo  {journal} {arXiv e-prints}\ ,\ \bibinfo {eid} {arXiv:2304.11911}} (\bibinfo {year} {2023})},\ \Eprint {https://arxiv.org/abs/2304.11911} {arXiv:2304.11911 [astro-ph.GA]} \BibitemShut {NoStop}%
\bibitem [{\citenamefont {{Tripodi}}\ \emph {et~al.}(2024)\citenamefont {{Tripodi}}, \citenamefont {{Martis}}, \citenamefont {{Markov}}, \citenamefont {{Brada{\v{c}}}}, \citenamefont {{Di Mascia}}, \citenamefont {{Cammelli}}, \citenamefont {{D'Eugenio}}, \citenamefont {{Willott}}, \citenamefont {{Curti}}, \citenamefont {{Bhatt}}, \citenamefont {{Gallerani}}, \citenamefont {{Rihtar{\v{s}}i{\v{c}}}}, \citenamefont {{Singh}}, \citenamefont {{Gaspar}}, \citenamefont {{Harshan}}, \citenamefont {{Jude{\v{z}}}}, \citenamefont {{Merida}}, \citenamefont {{Desprez}}, \citenamefont {{Sawicki}}, \citenamefont {{Goovaerts}}, \citenamefont {{Muzzin}}, \citenamefont {{Noirot}}, \citenamefont {{Sarrouh}}, \citenamefont {{Abraham}}, \citenamefont {{Asada}}, \citenamefont {{Brammer}}, \citenamefont {{Estrada Carpenter}}, \citenamefont {{Felicioni}}, \citenamefont {{Fujimoto}}, \citenamefont {{Iyer}}, \citenamefont {{Mowla}},\ and\ \citenamefont {{Strait}}}]{2024arXiv241204983T}%
  \BibitemOpen
  \bibfield  {author} {\bibinfo {author} {\bibfnamefont {R.}~\bibnamefont {{Tripodi}}}, \bibinfo {author} {\bibfnamefont {N.}~\bibnamefont {{Martis}}}, \bibinfo {author} {\bibfnamefont {V.}~\bibnamefont {{Markov}}}, \bibinfo {author} {\bibfnamefont {M.}~\bibnamefont {{Brada{\v{c}}}}}, \bibinfo {author} {\bibfnamefont {F.}~\bibnamefont {{Di Mascia}}}, \bibinfo {author} {\bibfnamefont {V.}~\bibnamefont {{Cammelli}}}, \bibinfo {author} {\bibfnamefont {F.}~\bibnamefont {{D'Eugenio}}}, \bibinfo {author} {\bibfnamefont {C.}~\bibnamefont {{Willott}}}, \bibinfo {author} {\bibfnamefont {M.}~\bibnamefont {{Curti}}}, \bibinfo {author} {\bibfnamefont {M.}~\bibnamefont {{Bhatt}}}, \bibinfo {author} {\bibfnamefont {S.}~\bibnamefont {{Gallerani}}}, \bibinfo {author} {\bibfnamefont {G.}~\bibnamefont {{Rihtar{\v{s}}i{\v{c}}}}}, \bibinfo {author} {\bibfnamefont {J.}~\bibnamefont {{Singh}}}, \bibinfo {author} {\bibfnamefont {G.}~\bibnamefont {{Gaspar}}}, \bibinfo {author} {\bibfnamefont {A.}~\bibnamefont {{Harshan}}}, \bibinfo
  {author} {\bibfnamefont {J.}~\bibnamefont {{Jude{\v{z}}}}}, \bibinfo {author} {\bibfnamefont {R.~M.}\ \bibnamefont {{Merida}}}, \bibinfo {author} {\bibfnamefont {G.}~\bibnamefont {{Desprez}}}, \bibinfo {author} {\bibfnamefont {M.}~\bibnamefont {{Sawicki}}}, \bibinfo {author} {\bibfnamefont {I.}~\bibnamefont {{Goovaerts}}}, \bibinfo {author} {\bibfnamefont {A.}~\bibnamefont {{Muzzin}}}, \bibinfo {author} {\bibfnamefont {G.}~\bibnamefont {{Noirot}}}, \bibinfo {author} {\bibfnamefont {G.~T.~E.}\ \bibnamefont {{Sarrouh}}}, \bibinfo {author} {\bibfnamefont {R.}~\bibnamefont {{Abraham}}}, \bibinfo {author} {\bibfnamefont {Y.}~\bibnamefont {{Asada}}}, \bibinfo {author} {\bibfnamefont {G.}~\bibnamefont {{Brammer}}}, \bibinfo {author} {\bibfnamefont {V.}~\bibnamefont {{Estrada Carpenter}}}, \bibinfo {author} {\bibfnamefont {G.}~\bibnamefont {{Felicioni}}}, \bibinfo {author} {\bibfnamefont {S.}~\bibnamefont {{Fujimoto}}}, \bibinfo {author} {\bibfnamefont {K.}~\bibnamefont {{Iyer}}}, \bibinfo {author} {\bibfnamefont
  {L.}~\bibnamefont {{Mowla}}},\ and\ \bibinfo {author} {\bibfnamefont {V.}~\bibnamefont {{Strait}}},\ }\href {https://doi.org/10.48550/arXiv.2412.04983} {\bibfield  {journal} {\bibinfo  {journal} {arXiv e-prints}\ ,\ \bibinfo {eid} {arXiv:2412.04983}} (\bibinfo {year} {2024})},\ \Eprint {https://arxiv.org/abs/2412.04983} {arXiv:2412.04983 [astro-ph.GA]} \BibitemShut {NoStop}%
\bibitem [{\citenamefont {{McCaffrey}}\ \emph {et~al.}(2023)\citenamefont {{McCaffrey}}, \citenamefont {{Hardin}}, \citenamefont {{Wise}},\ and\ \citenamefont {{Regan}}}]{2023OJAp....6E..47M}%
  \BibitemOpen
  \bibfield  {author} {\bibinfo {author} {\bibfnamefont {J.}~\bibnamefont {{McCaffrey}}}, \bibinfo {author} {\bibfnamefont {S.}~\bibnamefont {{Hardin}}}, \bibinfo {author} {\bibfnamefont {J.~H.}\ \bibnamefont {{Wise}}},\ and\ \bibinfo {author} {\bibfnamefont {J.~A.}\ \bibnamefont {{Regan}}},\ }\href {https://doi.org/10.21105/astro.2304.13755} {\bibfield  {journal} {\bibinfo  {journal} {The Open Journal of Astrophysics}\ }\textbf {\bibinfo {volume} {6}},\ \bibinfo {eid} {47} (\bibinfo {year} {2023})},\ \Eprint {https://arxiv.org/abs/2304.13755} {arXiv:2304.13755 [astro-ph.GA]} \BibitemShut {NoStop}%
\bibitem [{\citenamefont {{Harikane}}\ \emph {et~al.}(2024)\citenamefont {{Harikane}}, \citenamefont {{Inoue}}, \citenamefont {{Ellis}}, \citenamefont {{Ouchi}}, \citenamefont {{Nakazato}}, \citenamefont {{Yoshida}}, \citenamefont {{Ono}}, \citenamefont {{Sun}}, \citenamefont {{Sato}}, \citenamefont {{Ferrami}}, \citenamefont {{Fujimoto}}, \citenamefont {{Kashikawa}}, \citenamefont {{McLeod}}, \citenamefont {{Perez-Gonzalez}}, \citenamefont {{Sawicki}}, \citenamefont {{Sugahara}}, \citenamefont {{Xu}}, \citenamefont {{Yamanaka}}, \citenamefont {{Carnall}}, \citenamefont {{Cullen}}, \citenamefont {{Dunlop}}, \citenamefont {{Egami}}, \citenamefont {{Grogin}}, \citenamefont {{Isobe}}, \citenamefont {{Koekemoer}}, \citenamefont {{Laporte}}, \citenamefont {{Lee}}, \citenamefont {{Magee}}, \citenamefont {{Matsuo}}, \citenamefont {{Matsuoka}}, \citenamefont {{Mawatari}}, \citenamefont {{Nakajima}}, \citenamefont {{Nakane}}, \citenamefont {{Tamura}}, \citenamefont {{Umeda}},\ and\ \citenamefont
  {{Yanagisawa}}}]{2024arXiv240618352H}%
  \BibitemOpen
  \bibfield  {author} {\bibinfo {author} {\bibfnamefont {Y.}~\bibnamefont {{Harikane}}}, \bibinfo {author} {\bibfnamefont {A.~K.}\ \bibnamefont {{Inoue}}}, \bibinfo {author} {\bibfnamefont {R.~S.}\ \bibnamefont {{Ellis}}}, \bibinfo {author} {\bibfnamefont {M.}~\bibnamefont {{Ouchi}}}, \bibinfo {author} {\bibfnamefont {Y.}~\bibnamefont {{Nakazato}}}, \bibinfo {author} {\bibfnamefont {N.}~\bibnamefont {{Yoshida}}}, \bibinfo {author} {\bibfnamefont {Y.}~\bibnamefont {{Ono}}}, \bibinfo {author} {\bibfnamefont {F.}~\bibnamefont {{Sun}}}, \bibinfo {author} {\bibfnamefont {R.~A.}\ \bibnamefont {{Sato}}}, \bibinfo {author} {\bibfnamefont {G.}~\bibnamefont {{Ferrami}}}, \bibinfo {author} {\bibfnamefont {S.}~\bibnamefont {{Fujimoto}}}, \bibinfo {author} {\bibfnamefont {N.}~\bibnamefont {{Kashikawa}}}, \bibinfo {author} {\bibfnamefont {D.~J.}\ \bibnamefont {{McLeod}}}, \bibinfo {author} {\bibfnamefont {P.~G.}\ \bibnamefont {{Perez-Gonzalez}}}, \bibinfo {author} {\bibfnamefont {M.}~\bibnamefont {{Sawicki}}}, \bibinfo
  {author} {\bibfnamefont {Y.}~\bibnamefont {{Sugahara}}}, \bibinfo {author} {\bibfnamefont {Y.}~\bibnamefont {{Xu}}}, \bibinfo {author} {\bibfnamefont {S.}~\bibnamefont {{Yamanaka}}}, \bibinfo {author} {\bibfnamefont {A.~C.}\ \bibnamefont {{Carnall}}}, \bibinfo {author} {\bibfnamefont {F.}~\bibnamefont {{Cullen}}}, \bibinfo {author} {\bibfnamefont {J.~S.}\ \bibnamefont {{Dunlop}}}, \bibinfo {author} {\bibfnamefont {E.}~\bibnamefont {{Egami}}}, \bibinfo {author} {\bibfnamefont {N.}~\bibnamefont {{Grogin}}}, \bibinfo {author} {\bibfnamefont {Y.}~\bibnamefont {{Isobe}}}, \bibinfo {author} {\bibfnamefont {A.~M.}\ \bibnamefont {{Koekemoer}}}, \bibinfo {author} {\bibfnamefont {N.}~\bibnamefont {{Laporte}}}, \bibinfo {author} {\bibfnamefont {C.-H.}\ \bibnamefont {{Lee}}}, \bibinfo {author} {\bibfnamefont {D.}~\bibnamefont {{Magee}}}, \bibinfo {author} {\bibfnamefont {H.}~\bibnamefont {{Matsuo}}}, \bibinfo {author} {\bibfnamefont {Y.}~\bibnamefont {{Matsuoka}}}, \bibinfo {author} {\bibfnamefont {K.}~\bibnamefont
  {{Mawatari}}}, \bibinfo {author} {\bibfnamefont {K.}~\bibnamefont {{Nakajima}}}, \bibinfo {author} {\bibfnamefont {M.}~\bibnamefont {{Nakane}}}, \bibinfo {author} {\bibfnamefont {Y.}~\bibnamefont {{Tamura}}}, \bibinfo {author} {\bibfnamefont {H.}~\bibnamefont {{Umeda}}},\ and\ \bibinfo {author} {\bibfnamefont {H.}~\bibnamefont {{Yanagisawa}}},\ }\href {https://doi.org/10.48550/arXiv.2406.18352} {\bibfield  {journal} {\bibinfo  {journal} {arXiv e-prints}\ ,\ \bibinfo {eid} {arXiv:2406.18352}} (\bibinfo {year} {2024})},\ \Eprint {https://arxiv.org/abs/2406.18352} {arXiv:2406.18352 [astro-ph.GA]} \BibitemShut {NoStop}%
\bibitem [{\citenamefont {{Xiao}}\ \emph {et~al.}(2024)\citenamefont {{Xiao}}, \citenamefont {{Oesch}}, \citenamefont {{Elbaz}}, \citenamefont {{Bing}}, \citenamefont {{Nelson}}, \citenamefont {{Weibel}}, \citenamefont {{Illingworth}}, \citenamefont {{van Dokkum}}, \citenamefont {{Naidu}}, \citenamefont {{Daddi}}, \citenamefont {{Bouwens}}, \citenamefont {{Matthee}}, \citenamefont {{Wuyts}}, \citenamefont {{Chisholm}}, \citenamefont {{Brammer}}, \citenamefont {{Dickinson}}, \citenamefont {{Magnelli}}, \citenamefont {{Leroy}}, \citenamefont {{Schaerer}}, \citenamefont {{Herard-Demanche}}, \citenamefont {{Lim}}, \citenamefont {{Barrufet}}, \citenamefont {{Endsley}}, \citenamefont {{Fudamoto}}, \citenamefont {{G{\'o}mez-Guijarro}}, \citenamefont {{Gottumukkala}}, \citenamefont {{Labb{\'e}}}, \citenamefont {{Magee}}, \citenamefont {{Marchesini}}, \citenamefont {{Maseda}}, \citenamefont {{Qin}}, \citenamefont {{Reddy}}, \citenamefont {{Shapley}}, \citenamefont {{Shivaei}}, \citenamefont {{Shuntov}}, \citenamefont
  {{Stefanon}}, \citenamefont {{Whitaker}},\ and\ \citenamefont {{Wyithe}}}]{2024Natur.635..311X}%
  \BibitemOpen
  \bibfield  {author} {\bibinfo {author} {\bibfnamefont {M.}~\bibnamefont {{Xiao}}}, \bibinfo {author} {\bibfnamefont {P.~A.}\ \bibnamefont {{Oesch}}}, \bibinfo {author} {\bibfnamefont {D.}~\bibnamefont {{Elbaz}}}, \bibinfo {author} {\bibfnamefont {L.}~\bibnamefont {{Bing}}}, \bibinfo {author} {\bibfnamefont {E.~J.}\ \bibnamefont {{Nelson}}}, \bibinfo {author} {\bibfnamefont {A.}~\bibnamefont {{Weibel}}}, \bibinfo {author} {\bibfnamefont {G.~D.}\ \bibnamefont {{Illingworth}}}, \bibinfo {author} {\bibfnamefont {P.}~\bibnamefont {{van Dokkum}}}, \bibinfo {author} {\bibfnamefont {R.~P.}\ \bibnamefont {{Naidu}}}, \bibinfo {author} {\bibfnamefont {E.}~\bibnamefont {{Daddi}}}, \bibinfo {author} {\bibfnamefont {R.~J.}\ \bibnamefont {{Bouwens}}}, \bibinfo {author} {\bibfnamefont {J.}~\bibnamefont {{Matthee}}}, \bibinfo {author} {\bibfnamefont {S.}~\bibnamefont {{Wuyts}}}, \bibinfo {author} {\bibfnamefont {J.}~\bibnamefont {{Chisholm}}}, \bibinfo {author} {\bibfnamefont {G.}~\bibnamefont {{Brammer}}}, \bibinfo
  {author} {\bibfnamefont {M.}~\bibnamefont {{Dickinson}}}, \bibinfo {author} {\bibfnamefont {B.}~\bibnamefont {{Magnelli}}}, \bibinfo {author} {\bibfnamefont {L.}~\bibnamefont {{Leroy}}}, \bibinfo {author} {\bibfnamefont {D.}~\bibnamefont {{Schaerer}}}, \bibinfo {author} {\bibfnamefont {T.}~\bibnamefont {{Herard-Demanche}}}, \bibinfo {author} {\bibfnamefont {S.}~\bibnamefont {{Lim}}}, \bibinfo {author} {\bibfnamefont {L.}~\bibnamefont {{Barrufet}}}, \bibinfo {author} {\bibfnamefont {R.}~\bibnamefont {{Endsley}}}, \bibinfo {author} {\bibfnamefont {Y.}~\bibnamefont {{Fudamoto}}}, \bibinfo {author} {\bibfnamefont {C.}~\bibnamefont {{G{\'o}mez-Guijarro}}}, \bibinfo {author} {\bibfnamefont {R.}~\bibnamefont {{Gottumukkala}}}, \bibinfo {author} {\bibfnamefont {I.}~\bibnamefont {{Labb{\'e}}}}, \bibinfo {author} {\bibfnamefont {D.}~\bibnamefont {{Magee}}}, \bibinfo {author} {\bibfnamefont {D.}~\bibnamefont {{Marchesini}}}, \bibinfo {author} {\bibfnamefont {M.}~\bibnamefont {{Maseda}}}, \bibinfo {author}
  {\bibfnamefont {Y.}~\bibnamefont {{Qin}}}, \bibinfo {author} {\bibfnamefont {N.~A.}\ \bibnamefont {{Reddy}}}, \bibinfo {author} {\bibfnamefont {A.}~\bibnamefont {{Shapley}}}, \bibinfo {author} {\bibfnamefont {I.}~\bibnamefont {{Shivaei}}}, \bibinfo {author} {\bibfnamefont {M.}~\bibnamefont {{Shuntov}}}, \bibinfo {author} {\bibfnamefont {M.}~\bibnamefont {{Stefanon}}}, \bibinfo {author} {\bibfnamefont {K.~E.}\ \bibnamefont {{Whitaker}}},\ and\ \bibinfo {author} {\bibfnamefont {J.~S.~B.}\ \bibnamefont {{Wyithe}}},\ }\href {https://doi.org/10.1038/s41586-024-08094-5} {\bibfield  {journal} {\bibinfo  {journal} {\nat}\ }\textbf {\bibinfo {volume} {635}},\ \bibinfo {pages} {311} (\bibinfo {year} {2024})},\ \Eprint {https://arxiv.org/abs/2309.02492} {arXiv:2309.02492 [astro-ph.GA]} \BibitemShut {NoStop}%
\bibitem [{\citenamefont {{Harvey}}\ \emph {et~al.}(2025)\citenamefont {{Harvey}}, \citenamefont {{Conselice}}, \citenamefont {{Adams}}, \citenamefont {{Austin}}, \citenamefont {{Juod{\v{z}}balis}}, \citenamefont {{Trussler}}, \citenamefont {{Li}}, \citenamefont {{Ormerod}}, \citenamefont {{Ferreira}}, \citenamefont {{Lovell}}, \citenamefont {{Duan}}, \citenamefont {{Westcott}}, \citenamefont {{Harris}}, \citenamefont {{Bhatawdekar}}, \citenamefont {{Coe}}, \citenamefont {{Cohen}}, \citenamefont {{Caruana}}, \citenamefont {{Cheng}}, \citenamefont {{Driver}}, \citenamefont {{Frye}}, \citenamefont {{Furtak}}, \citenamefont {{Grogin}}, \citenamefont {{Hathi}}, \citenamefont {{Holwerda}}, \citenamefont {{Jansen}}, \citenamefont {{Koekemoer}}, \citenamefont {{Marshall}}, \citenamefont {{Nonino}}, \citenamefont {{Vijayan}}, \citenamefont {{Wilkins}}, \citenamefont {{Windhorst}}, \citenamefont {{Willmer}}, \citenamefont {{Yan}},\ and\ \citenamefont {{Zitrin}}}]{2025ApJ...978...89H}%
  \BibitemOpen
  \bibfield  {author} {\bibinfo {author} {\bibfnamefont {T.}~\bibnamefont {{Harvey}}}, \bibinfo {author} {\bibfnamefont {C.~J.}\ \bibnamefont {{Conselice}}}, \bibinfo {author} {\bibfnamefont {N.~J.}\ \bibnamefont {{Adams}}}, \bibinfo {author} {\bibfnamefont {D.}~\bibnamefont {{Austin}}}, \bibinfo {author} {\bibfnamefont {I.}~\bibnamefont {{Juod{\v{z}}balis}}}, \bibinfo {author} {\bibfnamefont {J.}~\bibnamefont {{Trussler}}}, \bibinfo {author} {\bibfnamefont {Q.}~\bibnamefont {{Li}}}, \bibinfo {author} {\bibfnamefont {K.}~\bibnamefont {{Ormerod}}}, \bibinfo {author} {\bibfnamefont {L.}~\bibnamefont {{Ferreira}}}, \bibinfo {author} {\bibfnamefont {C.~C.}\ \bibnamefont {{Lovell}}}, \bibinfo {author} {\bibfnamefont {Q.}~\bibnamefont {{Duan}}}, \bibinfo {author} {\bibfnamefont {L.}~\bibnamefont {{Westcott}}}, \bibinfo {author} {\bibfnamefont {H.}~\bibnamefont {{Harris}}}, \bibinfo {author} {\bibfnamefont {R.}~\bibnamefont {{Bhatawdekar}}}, \bibinfo {author} {\bibfnamefont {D.}~\bibnamefont {{Coe}}}, \bibinfo
  {author} {\bibfnamefont {S.~H.}\ \bibnamefont {{Cohen}}}, \bibinfo {author} {\bibfnamefont {J.}~\bibnamefont {{Caruana}}}, \bibinfo {author} {\bibfnamefont {C.}~\bibnamefont {{Cheng}}}, \bibinfo {author} {\bibfnamefont {S.~P.}\ \bibnamefont {{Driver}}}, \bibinfo {author} {\bibfnamefont {B.}~\bibnamefont {{Frye}}}, \bibinfo {author} {\bibfnamefont {L.~J.}\ \bibnamefont {{Furtak}}}, \bibinfo {author} {\bibfnamefont {N.~A.}\ \bibnamefont {{Grogin}}}, \bibinfo {author} {\bibfnamefont {N.~P.}\ \bibnamefont {{Hathi}}}, \bibinfo {author} {\bibfnamefont {B.~W.}\ \bibnamefont {{Holwerda}}}, \bibinfo {author} {\bibfnamefont {R.~A.}\ \bibnamefont {{Jansen}}}, \bibinfo {author} {\bibfnamefont {A.~M.}\ \bibnamefont {{Koekemoer}}}, \bibinfo {author} {\bibfnamefont {M.~A.}\ \bibnamefont {{Marshall}}}, \bibinfo {author} {\bibfnamefont {M.}~\bibnamefont {{Nonino}}}, \bibinfo {author} {\bibfnamefont {A.~P.}\ \bibnamefont {{Vijayan}}}, \bibinfo {author} {\bibfnamefont {S.~M.}\ \bibnamefont {{Wilkins}}}, \bibinfo {author}
  {\bibfnamefont {R.}~\bibnamefont {{Windhorst}}}, \bibinfo {author} {\bibfnamefont {C.~N.~A.}\ \bibnamefont {{Willmer}}}, \bibinfo {author} {\bibfnamefont {H.}~\bibnamefont {{Yan}}},\ and\ \bibinfo {author} {\bibfnamefont {A.}~\bibnamefont {{Zitrin}}},\ }\href {https://doi.org/10.3847/1538-4357/ad8c29} {\bibfield  {journal} {\bibinfo  {journal} {\apj}\ }\textbf {\bibinfo {volume} {978}},\ \bibinfo {eid} {89} (\bibinfo {year} {2025})},\ \Eprint {https://arxiv.org/abs/2403.03908} {arXiv:2403.03908 [astro-ph.GA]} \BibitemShut {NoStop}%
\bibitem [{\citenamefont {{Shen}}\ \emph {et~al.}(2023)\citenamefont {{Shen}}, \citenamefont {{Vogelsberger}}, \citenamefont {{Boylan-Kolchin}}, \citenamefont {{Tacchella}},\ and\ \citenamefont {{Kannan}}}]{2023MNRAS.525.3254S}%
  \BibitemOpen
  \bibfield  {author} {\bibinfo {author} {\bibfnamefont {X.}~\bibnamefont {{Shen}}}, \bibinfo {author} {\bibfnamefont {M.}~\bibnamefont {{Vogelsberger}}}, \bibinfo {author} {\bibfnamefont {M.}~\bibnamefont {{Boylan-Kolchin}}}, \bibinfo {author} {\bibfnamefont {S.}~\bibnamefont {{Tacchella}}},\ and\ \bibinfo {author} {\bibfnamefont {R.}~\bibnamefont {{Kannan}}},\ }\href {https://doi.org/10.1093/mnras/stad2508} {\bibfield  {journal} {\bibinfo  {journal} {\mnras}\ }\textbf {\bibinfo {volume} {525}},\ \bibinfo {pages} {3254} (\bibinfo {year} {2023})},\ \Eprint {https://arxiv.org/abs/2305.05679} {arXiv:2305.05679 [astro-ph.GA]} \BibitemShut {NoStop}%
\bibitem [{\citenamefont {{Lu}}\ \emph {et~al.}(2025)\citenamefont {{Lu}}, \citenamefont {{Frenk}}, \citenamefont {{Bose}}, \citenamefont {{Lacey}}, \citenamefont {{Cole}}, \citenamefont {{Baugh}},\ and\ \citenamefont {{Helly}}}]{2025MNRAS.536.1018L}%
  \BibitemOpen
  \bibfield  {author} {\bibinfo {author} {\bibfnamefont {S.}~\bibnamefont {{Lu}}}, \bibinfo {author} {\bibfnamefont {C.~S.}\ \bibnamefont {{Frenk}}}, \bibinfo {author} {\bibfnamefont {S.}~\bibnamefont {{Bose}}}, \bibinfo {author} {\bibfnamefont {C.~G.}\ \bibnamefont {{Lacey}}}, \bibinfo {author} {\bibfnamefont {S.}~\bibnamefont {{Cole}}}, \bibinfo {author} {\bibfnamefont {C.~M.}\ \bibnamefont {{Baugh}}},\ and\ \bibinfo {author} {\bibfnamefont {J.~C.}\ \bibnamefont {{Helly}}},\ }\href {https://doi.org/10.1093/mnras/stae2646} {\bibfield  {journal} {\bibinfo  {journal} {\mnras}\ }\textbf {\bibinfo {volume} {536}},\ \bibinfo {pages} {1018} (\bibinfo {year} {2025})},\ \Eprint {https://arxiv.org/abs/2406.02672} {arXiv:2406.02672 [astro-ph.GA]} \BibitemShut {NoStop}%
\bibitem [{\citenamefont {{Mirocha}}\ and\ \citenamefont {{Furlanetto}}(2023)}]{Mirocha2023MNRAS.519..843M}%
  \BibitemOpen
  \bibfield  {author} {\bibinfo {author} {\bibfnamefont {J.}~\bibnamefont {{Mirocha}}}\ and\ \bibinfo {author} {\bibfnamefont {S.~R.}\ \bibnamefont {{Furlanetto}}},\ }\href {https://doi.org/10.1093/mnras/stac3578} {\bibfield  {journal} {\bibinfo  {journal} {\mnras}\ }\textbf {\bibinfo {volume} {519}},\ \bibinfo {pages} {843} (\bibinfo {year} {2023})},\ \Eprint {https://arxiv.org/abs/2208.12826} {arXiv:2208.12826 [astro-ph.GA]} \BibitemShut {NoStop}%
\bibitem [{\citenamefont {{Ferrara}}\ \emph {et~al.}(2023)\citenamefont {{Ferrara}}, \citenamefont {{Pallottini}},\ and\ \citenamefont {{Dayal}}}]{ferrara2023MNRAS.522.3986F}%
  \BibitemOpen
  \bibfield  {author} {\bibinfo {author} {\bibfnamefont {A.}~\bibnamefont {{Ferrara}}}, \bibinfo {author} {\bibfnamefont {A.}~\bibnamefont {{Pallottini}}},\ and\ \bibinfo {author} {\bibfnamefont {P.}~\bibnamefont {{Dayal}}},\ }\href {https://doi.org/10.1093/mnras/stad1095} {\bibfield  {journal} {\bibinfo  {journal} {\mnras}\ }\textbf {\bibinfo {volume} {522}},\ \bibinfo {pages} {3986} (\bibinfo {year} {2023})},\ \Eprint {https://arxiv.org/abs/2208.00720} {arXiv:2208.00720 [astro-ph.GA]} \BibitemShut {NoStop}%
\bibitem [{\citenamefont {{Mason}}\ \emph {et~al.}(2023)\citenamefont {{Mason}}, \citenamefont {{Trenti}},\ and\ \citenamefont {{Treu}}}]{Mason2023MNRAS.521..497M}%
  \BibitemOpen
  \bibfield  {author} {\bibinfo {author} {\bibfnamefont {C.~A.}\ \bibnamefont {{Mason}}}, \bibinfo {author} {\bibfnamefont {M.}~\bibnamefont {{Trenti}}},\ and\ \bibinfo {author} {\bibfnamefont {T.}~\bibnamefont {{Treu}}},\ }\href {https://doi.org/10.1093/mnras/stad035} {\bibfield  {journal} {\bibinfo  {journal} {\mnras}\ }\textbf {\bibinfo {volume} {521}},\ \bibinfo {pages} {497} (\bibinfo {year} {2023})},\ \Eprint {https://arxiv.org/abs/2207.14808} {arXiv:2207.14808 [astro-ph.GA]} \BibitemShut {NoStop}%
\bibitem [{\citenamefont {{Heather}}\ \emph {et~al.}(2024)\citenamefont {{Heather}}, \citenamefont {{Chantavat}}, \citenamefont {{Chongchitnan}},\ and\ \citenamefont {{Silk}}}]{Heather}%
  \BibitemOpen
  \bibfield  {author} {\bibinfo {author} {\bibfnamefont {C.}~\bibnamefont {{Heather}}}, \bibinfo {author} {\bibfnamefont {T.}~\bibnamefont {{Chantavat}}}, \bibinfo {author} {\bibfnamefont {S.}~\bibnamefont {{Chongchitnan}}},\ and\ \bibinfo {author} {\bibfnamefont {J.}~\bibnamefont {{Silk}}},\ }\href {https://doi.org/10.1093/mnras/stae2051} {\bibfield  {journal} {\bibinfo  {journal} {Mon. Not. Roy. Astron. Soc.}\ }\textbf {\bibinfo {volume} {534}},\ \bibinfo {pages} {173} (\bibinfo {year} {2024})},\ \Eprint {https://arxiv.org/abs/2404.11567} {arXiv:2404.11567 [astro-ph.GA]} \BibitemShut {NoStop}%
\bibitem [{\citenamefont {{Carnall}}\ \emph {et~al.}(2024)\citenamefont {{Carnall}}, \citenamefont {{Cullen}}, \citenamefont {{McLure}}, \citenamefont {{McLeod}}, \citenamefont {{Begley}}, \citenamefont {{Donnan}}, \citenamefont {{Dunlop}}, \citenamefont {{Shapley}}, \citenamefont {{Rowlands}}, \citenamefont {{Almaini}}, \citenamefont {{Arellano-C{\'o}rdova}}, \citenamefont {{Barrufet}}, \citenamefont {{Cimatti}}, \citenamefont {{Ellis}}, \citenamefont {{Grogin}}, \citenamefont {{Hamadouche}}, \citenamefont {{Illingworth}}, \citenamefont {{Koekemoer}}, \citenamefont {{Leung}}, \citenamefont {{Lovell}}, \citenamefont {{P{\'e}rez-Gonz{\'a}lez}}, \citenamefont {{Santini}}, \citenamefont {{Stanton}},\ and\ \citenamefont {{Wild}}}]{Carnall}%
  \BibitemOpen
  \bibfield  {author} {\bibinfo {author} {\bibfnamefont {A.~C.}\ \bibnamefont {{Carnall}}}, \bibinfo {author} {\bibfnamefont {F.}~\bibnamefont {{Cullen}}}, \bibinfo {author} {\bibfnamefont {R.~J.}\ \bibnamefont {{McLure}}}, \bibinfo {author} {\bibfnamefont {D.~J.}\ \bibnamefont {{McLeod}}}, \bibinfo {author} {\bibfnamefont {R.}~\bibnamefont {{Begley}}}, \bibinfo {author} {\bibfnamefont {C.~T.}\ \bibnamefont {{Donnan}}}, \bibinfo {author} {\bibfnamefont {J.~S.}\ \bibnamefont {{Dunlop}}}, \bibinfo {author} {\bibfnamefont {A.~E.}\ \bibnamefont {{Shapley}}}, \bibinfo {author} {\bibfnamefont {K.}~\bibnamefont {{Rowlands}}}, \bibinfo {author} {\bibfnamefont {O.}~\bibnamefont {{Almaini}}}, \bibinfo {author} {\bibfnamefont {K.~Z.}\ \bibnamefont {{Arellano-C{\'o}rdova}}}, \bibinfo {author} {\bibfnamefont {L.}~\bibnamefont {{Barrufet}}}, \bibinfo {author} {\bibfnamefont {A.}~\bibnamefont {{Cimatti}}}, \bibinfo {author} {\bibfnamefont {R.~S.}\ \bibnamefont {{Ellis}}}, \bibinfo {author} {\bibfnamefont {N.~A.}\
  \bibnamefont {{Grogin}}}, \bibinfo {author} {\bibfnamefont {M.~L.}\ \bibnamefont {{Hamadouche}}}, \bibinfo {author} {\bibfnamefont {G.~D.}\ \bibnamefont {{Illingworth}}}, \bibinfo {author} {\bibfnamefont {A.~M.}\ \bibnamefont {{Koekemoer}}}, \bibinfo {author} {\bibfnamefont {H.~H.}\ \bibnamefont {{Leung}}}, \bibinfo {author} {\bibfnamefont {C.~C.}\ \bibnamefont {{Lovell}}}, \bibinfo {author} {\bibfnamefont {P.~G.}\ \bibnamefont {{P{\'e}rez-Gonz{\'a}lez}}}, \bibinfo {author} {\bibfnamefont {P.}~\bibnamefont {{Santini}}}, \bibinfo {author} {\bibfnamefont {T.~M.}\ \bibnamefont {{Stanton}}},\ and\ \bibinfo {author} {\bibfnamefont {V.}~\bibnamefont {{Wild}}},\ }\href {https://doi.org/10.1093/mnras/stae2092} {\bibfield  {journal} {\bibinfo  {journal} {Mon. Not. Roy. Astron. Soc.}\ }\textbf {\bibinfo {volume} {534}},\ \bibinfo {pages} {325} (\bibinfo {year} {2024})},\ \Eprint {https://arxiv.org/abs/2405.02242} {arXiv:2405.02242 [astro-ph.GA]} \BibitemShut {NoStop}%
\bibitem [{\citenamefont {Lovell}\ \emph {et~al.}(2022)\citenamefont {Lovell}, \citenamefont {Harrison}, \citenamefont {Harikane}, \citenamefont {Tacchella},\ and\ \citenamefont {Wilkins}}]{Lovell:2022bhx}%
  \BibitemOpen
  \bibfield  {author} {\bibinfo {author} {\bibfnamefont {C.~C.}\ \bibnamefont {Lovell}}, \bibinfo {author} {\bibfnamefont {I.}~\bibnamefont {Harrison}}, \bibinfo {author} {\bibfnamefont {Y.}~\bibnamefont {Harikane}}, \bibinfo {author} {\bibfnamefont {S.}~\bibnamefont {Tacchella}},\ and\ \bibinfo {author} {\bibfnamefont {S.~M.}\ \bibnamefont {Wilkins}},\ }\href {https://doi.org/10.1093/mnras/stac3224} {\bibfield  {journal} {\bibinfo  {journal} {Mon. Not. Roy. Astron. Soc.}\ }\textbf {\bibinfo {volume} {518}},\ \bibinfo {pages} {2511} (\bibinfo {year} {2022})},\ \Eprint {https://arxiv.org/abs/2208.10479} {arXiv:2208.10479 [astro-ph.GA]} \BibitemShut {NoStop}%
\bibitem [{\citenamefont {{Boylan-Kolchin}}(2023)}]{2023NatAs...7..731B}%
  \BibitemOpen
  \bibfield  {author} {\bibinfo {author} {\bibfnamefont {M.}~\bibnamefont {{Boylan-Kolchin}}},\ }\href {https://doi.org/10.1038/s41550-023-01937-7} {\bibfield  {journal} {\bibinfo  {journal} {Nature Astronomy}\ }\textbf {\bibinfo {volume} {7}},\ \bibinfo {pages} {731} (\bibinfo {year} {2023})},\ \Eprint {https://arxiv.org/abs/2208.01611} {arXiv:2208.01611 [astro-ph.CO]} \BibitemShut {NoStop}%
\bibitem [{\citenamefont {{Shen}}\ \emph {et~al.}(2024)\citenamefont {{Shen}}, \citenamefont {{Vogelsberger}}, \citenamefont {{Boylan-Kolchin}}, \citenamefont {{Tacchella}},\ and\ \citenamefont {{Naidu}}}]{2024MNRAS.533.3923S}%
  \BibitemOpen
  \bibfield  {author} {\bibinfo {author} {\bibfnamefont {X.}~\bibnamefont {{Shen}}}, \bibinfo {author} {\bibfnamefont {M.}~\bibnamefont {{Vogelsberger}}}, \bibinfo {author} {\bibfnamefont {M.}~\bibnamefont {{Boylan-Kolchin}}}, \bibinfo {author} {\bibfnamefont {S.}~\bibnamefont {{Tacchella}}},\ and\ \bibinfo {author} {\bibfnamefont {R.~P.}\ \bibnamefont {{Naidu}}},\ }\href {https://doi.org/10.1093/mnras/stae1932} {\bibfield  {journal} {\bibinfo  {journal} {\mnras}\ }\textbf {\bibinfo {volume} {533}},\ \bibinfo {pages} {3923} (\bibinfo {year} {2024})},\ \Eprint {https://arxiv.org/abs/2406.15548} {arXiv:2406.15548 [astro-ph.GA]} \BibitemShut {NoStop}%
\bibitem [{\citenamefont {Linder}(2005)}]{Linder:2005in}%
  \BibitemOpen
  \bibfield  {author} {\bibinfo {author} {\bibfnamefont {E.~V.}\ \bibnamefont {Linder}},\ }\href {https://doi.org/10.1103/PhysRevD.72.043529} {\bibfield  {journal} {\bibinfo  {journal} {Phys. Rev. D}\ }\textbf {\bibinfo {volume} {72}},\ \bibinfo {pages} {043529} (\bibinfo {year} {2005})},\ \Eprint {https://arxiv.org/abs/astro-ph/0507263} {arXiv:astro-ph/0507263} \BibitemShut {NoStop}%
\bibitem [{\citenamefont {Linder}\ and\ \citenamefont {Cahn}(2007)}]{Linder:2007hg}%
  \BibitemOpen
  \bibfield  {author} {\bibinfo {author} {\bibfnamefont {E.~V.}\ \bibnamefont {Linder}}\ and\ \bibinfo {author} {\bibfnamefont {R.~N.}\ \bibnamefont {Cahn}},\ }\href {https://doi.org/10.1016/j.astropartphys.2007.09.003} {\bibfield  {journal} {\bibinfo  {journal} {Astropart. Phys.}\ }\textbf {\bibinfo {volume} {28}},\ \bibinfo {pages} {481} (\bibinfo {year} {2007})},\ \Eprint {https://arxiv.org/abs/astro-ph/0701317} {arXiv:astro-ph/0701317} \BibitemShut {NoStop}%
\bibitem [{\citenamefont {Gumbel}(1958)}]{Gumbel+1958}%
  \BibitemOpen
  \bibfield  {author} {\bibinfo {author} {\bibfnamefont {E.~J.}\ \bibnamefont {Gumbel}},\ }\href {https://doi.org/doi:10.7312/gumb92958} {\emph {\bibinfo {title} {Statistics of Extremes}}}\ (\bibinfo  {publisher} {Columbia University Press},\ \bibinfo {address} {New York Chichester, West Sussex},\ \bibinfo {year} {1958})\BibitemShut {NoStop}%
\bibitem [{\citenamefont {Aghanim}\ \emph {et~al.}(2020)\citenamefont {Aghanim} \emph {et~al.}}]{Planck:2018vyg}%
  \BibitemOpen
  \bibfield  {author} {\bibinfo {author} {\bibfnamefont {N.}~\bibnamefont {Aghanim}} \emph {et~al.} (\bibinfo {collaboration} {Planck}),\ }\href {https://doi.org/10.1051/0004-6361/201833910} {\bibfield  {journal} {\bibinfo  {journal} {Astron. Astrophys.}\ }\textbf {\bibinfo {volume} {641}},\ \bibinfo {pages} {A6} (\bibinfo {year} {2020})},\ \bibinfo {note} {[Erratum: Astron.Astrophys. 652, C4 (2021)]},\ \Eprint {https://arxiv.org/abs/1807.06209} {arXiv:1807.06209 [astro-ph.CO]} \BibitemShut {NoStop}%
\bibitem [{\citenamefont {Sheth}\ and\ \citenamefont {Tormen}(2002)}]{Sheth:2001dp}%
  \BibitemOpen
  \bibfield  {author} {\bibinfo {author} {\bibfnamefont {R.~K.}\ \bibnamefont {Sheth}}\ and\ \bibinfo {author} {\bibfnamefont {G.}~\bibnamefont {Tormen}},\ }\href {https://doi.org/10.1046/j.1365-8711.2002.04950.x} {\bibfield  {journal} {\bibinfo  {journal} {Mon. Not. Roy. Astron. Soc.}\ }\textbf {\bibinfo {volume} {329}},\ \bibinfo {pages} {61} (\bibinfo {year} {2002})},\ \Eprint {https://arxiv.org/abs/astro-ph/0105113} {arXiv:astro-ph/0105113} \BibitemShut {NoStop}%
\bibitem [{\citenamefont {Jenkins}\ \emph {et~al.}(2001)\citenamefont {Jenkins}, \citenamefont {Frenk}, \citenamefont {White}, \citenamefont {Colberg}, \citenamefont {Cole}, \citenamefont {Evrard}, \citenamefont {Couchman},\ and\ \citenamefont {Yoshida}}]{Jenkins:2000bv}%
  \BibitemOpen
  \bibfield  {author} {\bibinfo {author} {\bibfnamefont {A.}~\bibnamefont {Jenkins}}, \bibinfo {author} {\bibfnamefont {C.~S.}\ \bibnamefont {Frenk}}, \bibinfo {author} {\bibfnamefont {S.~D.~M.}\ \bibnamefont {White}}, \bibinfo {author} {\bibfnamefont {J.~M.}\ \bibnamefont {Colberg}}, \bibinfo {author} {\bibfnamefont {S.}~\bibnamefont {Cole}}, \bibinfo {author} {\bibfnamefont {A.~E.}\ \bibnamefont {Evrard}}, \bibinfo {author} {\bibfnamefont {H.~M.~P.}\ \bibnamefont {Couchman}},\ and\ \bibinfo {author} {\bibfnamefont {N.}~\bibnamefont {Yoshida}},\ }\href {https://doi.org/10.1046/j.1365-8711.2001.04029.x} {\bibfield  {journal} {\bibinfo  {journal} {Mon. Not. Roy. Astron. Soc.}\ }\textbf {\bibinfo {volume} {321}},\ \bibinfo {pages} {372} (\bibinfo {year} {2001})},\ \Eprint {https://arxiv.org/abs/astro-ph/0005260} {arXiv:astro-ph/0005260} \BibitemShut {NoStop}%
\bibitem [{\citenamefont {Tinker}\ \emph {et~al.}(2008)\citenamefont {Tinker}, \citenamefont {Kravtsov}, \citenamefont {Klypin}, \citenamefont {Abazajian}, \citenamefont {Warren}, \citenamefont {Yepes}, \citenamefont {Gottlober},\ and\ \citenamefont {Holz}}]{Tinker:2008ff}%
  \BibitemOpen
  \bibfield  {author} {\bibinfo {author} {\bibfnamefont {J.~L.}\ \bibnamefont {Tinker}}, \bibinfo {author} {\bibfnamefont {A.~V.}\ \bibnamefont {Kravtsov}}, \bibinfo {author} {\bibfnamefont {A.}~\bibnamefont {Klypin}}, \bibinfo {author} {\bibfnamefont {K.}~\bibnamefont {Abazajian}}, \bibinfo {author} {\bibfnamefont {M.~S.}\ \bibnamefont {Warren}}, \bibinfo {author} {\bibfnamefont {G.}~\bibnamefont {Yepes}}, \bibinfo {author} {\bibfnamefont {S.}~\bibnamefont {Gottlober}},\ and\ \bibinfo {author} {\bibfnamefont {D.~E.}\ \bibnamefont {Holz}},\ }\href {https://doi.org/10.1086/591439} {\bibfield  {journal} {\bibinfo  {journal} {Astrophys. J.}\ }\textbf {\bibinfo {volume} {688}},\ \bibinfo {pages} {709} (\bibinfo {year} {2008})},\ \Eprint {https://arxiv.org/abs/0803.2706} {arXiv:0803.2706 [astro-ph]} \BibitemShut {NoStop}%
\bibitem [{\citenamefont {Reed}\ \emph {et~al.}(2007)\citenamefont {Reed}, \citenamefont {Bower}, \citenamefont {Frenk}, \citenamefont {Jenkins},\ and\ \citenamefont {Theuns}}]{Reed:2006rw}%
  \BibitemOpen
  \bibfield  {author} {\bibinfo {author} {\bibfnamefont {D.}~\bibnamefont {Reed}}, \bibinfo {author} {\bibfnamefont {R.}~\bibnamefont {Bower}}, \bibinfo {author} {\bibfnamefont {C.}~\bibnamefont {Frenk}}, \bibinfo {author} {\bibfnamefont {A.}~\bibnamefont {Jenkins}},\ and\ \bibinfo {author} {\bibfnamefont {T.}~\bibnamefont {Theuns}},\ }\href {https://doi.org/10.1111/j.1365-2966.2006.11204.x} {\bibfield  {journal} {\bibinfo  {journal} {Mon. Not. Roy. Astron. Soc.}\ }\textbf {\bibinfo {volume} {374}},\ \bibinfo {pages} {2} (\bibinfo {year} {2007})},\ \Eprint {https://arxiv.org/abs/astro-ph/0607150} {arXiv:astro-ph/0607150} \BibitemShut {NoStop}%
\bibitem [{\citenamefont {Warren}\ \emph {et~al.}(2006)\citenamefont {Warren}, \citenamefont {Abazajian}, \citenamefont {Holz},\ and\ \citenamefont {Teodoro}}]{Warren:2005ey}%
  \BibitemOpen
  \bibfield  {author} {\bibinfo {author} {\bibfnamefont {M.~S.}\ \bibnamefont {Warren}}, \bibinfo {author} {\bibfnamefont {K.}~\bibnamefont {Abazajian}}, \bibinfo {author} {\bibfnamefont {D.~E.}\ \bibnamefont {Holz}},\ and\ \bibinfo {author} {\bibfnamefont {L.}~\bibnamefont {Teodoro}},\ }\href {https://doi.org/10.1086/504962} {\bibfield  {journal} {\bibinfo  {journal} {Astrophys. J.}\ }\textbf {\bibinfo {volume} {646}},\ \bibinfo {pages} {881} (\bibinfo {year} {2006})},\ \Eprint {https://arxiv.org/abs/astro-ph/0506395} {arXiv:astro-ph/0506395} \BibitemShut {NoStop}%
\bibitem [{\citenamefont {Lukic}\ \emph {et~al.}(2007)\citenamefont {Lukic}, \citenamefont {Heitmann}, \citenamefont {Habib}, \citenamefont {Bashinsky},\ and\ \citenamefont {Ricker}}]{Lukic:2007fc}%
  \BibitemOpen
  \bibfield  {author} {\bibinfo {author} {\bibfnamefont {Z.}~\bibnamefont {Lukic}}, \bibinfo {author} {\bibfnamefont {K.}~\bibnamefont {Heitmann}}, \bibinfo {author} {\bibfnamefont {S.}~\bibnamefont {Habib}}, \bibinfo {author} {\bibfnamefont {S.}~\bibnamefont {Bashinsky}},\ and\ \bibinfo {author} {\bibfnamefont {P.~M.}\ \bibnamefont {Ricker}},\ }\href {https://doi.org/10.1086/523083} {\bibfield  {journal} {\bibinfo  {journal} {Astrophys. J.}\ }\textbf {\bibinfo {volume} {671}},\ \bibinfo {pages} {1160} (\bibinfo {year} {2007})},\ \Eprint {https://arxiv.org/abs/astro-ph/0702360} {arXiv:astro-ph/0702360} \BibitemShut {NoStop}%
\bibitem [{\citenamefont {{Eddington}}(1913)}]{Eddington}%
  \BibitemOpen
  \bibfield  {author} {\bibinfo {author} {\bibfnamefont {A.~S.}\ \bibnamefont {{Eddington}}},\ }\href {https://doi.org/10.1093/mnras/73.5.359} {\bibfield  {journal} {\bibinfo  {journal} {Mon. Not. Roy. Astron. Soc.}\ }\textbf {\bibinfo {volume} {73}},\ \bibinfo {pages} {359} (\bibinfo {year} {1913})}\BibitemShut {NoStop}%
\bibitem [{\citenamefont {Stanek}\ \emph {et~al.}(2006)\citenamefont {Stanek}, \citenamefont {Evrard}, \citenamefont {Bohringer}, \citenamefont {Schuecker},\ and\ \citenamefont {Nord}}]{Stanek:2006tu}%
  \BibitemOpen
  \bibfield  {author} {\bibinfo {author} {\bibfnamefont {R.}~\bibnamefont {Stanek}}, \bibinfo {author} {\bibfnamefont {A.~E.}\ \bibnamefont {Evrard}}, \bibinfo {author} {\bibfnamefont {H.~B.}\ \bibnamefont {Bohringer}}, \bibinfo {author} {\bibfnamefont {P.}~\bibnamefont {Schuecker}},\ and\ \bibinfo {author} {\bibfnamefont {B.}~\bibnamefont {Nord}},\ }\href {https://doi.org/10.1086/506248} {\bibfield  {journal} {\bibinfo  {journal} {Astrophys. J.}\ }\textbf {\bibinfo {volume} {648}},\ \bibinfo {pages} {956} (\bibinfo {year} {2006})},\ \Eprint {https://arxiv.org/abs/astro-ph/0602324} {arXiv:astro-ph/0602324} \BibitemShut {NoStop}%
\bibitem [{\citenamefont {Mortonson}\ \emph {et~al.}(2011)\citenamefont {Mortonson}, \citenamefont {Hu},\ and\ \citenamefont {Huterer}}]{Mortonson:2010mj}%
  \BibitemOpen
  \bibfield  {author} {\bibinfo {author} {\bibfnamefont {M.~J.}\ \bibnamefont {Mortonson}}, \bibinfo {author} {\bibfnamefont {W.}~\bibnamefont {Hu}},\ and\ \bibinfo {author} {\bibfnamefont {D.}~\bibnamefont {Huterer}},\ }\href {https://doi.org/10.1103/PhysRevD.83.023015} {\bibfield  {journal} {\bibinfo  {journal} {Phys. Rev. D}\ }\textbf {\bibinfo {volume} {83}},\ \bibinfo {pages} {023015} (\bibinfo {year} {2011})},\ \Eprint {https://arxiv.org/abs/1011.0004} {arXiv:1011.0004 [astro-ph.CO]} \BibitemShut {NoStop}%
\bibitem [{\citenamefont {Nguyen}\ \emph {et~al.}(2023)\citenamefont {Nguyen}, \citenamefont {Huterer},\ and\ \citenamefont {Wen}}]{Nguyen:2023fip}%
  \BibitemOpen
  \bibfield  {author} {\bibinfo {author} {\bibfnamefont {N.-M.}\ \bibnamefont {Nguyen}}, \bibinfo {author} {\bibfnamefont {D.}~\bibnamefont {Huterer}},\ and\ \bibinfo {author} {\bibfnamefont {Y.}~\bibnamefont {Wen}},\ }\href {https://doi.org/10.1103/PhysRevLett.131.111001} {\bibfield  {journal} {\bibinfo  {journal} {Phys. Rev. Lett.}\ }\textbf {\bibinfo {volume} {131}},\ \bibinfo {pages} {111001} (\bibinfo {year} {2023})},\ \Eprint {https://arxiv.org/abs/2302.01331} {arXiv:2302.01331 [astro-ph.CO]} \BibitemShut {NoStop}%
\bibitem [{\citenamefont {{Press}}\ and\ \citenamefont {{Schechter}}(1974)}]{1974ApJ...187..425P}%
  \BibitemOpen
  \bibfield  {author} {\bibinfo {author} {\bibfnamefont {W.~H.}\ \bibnamefont {{Press}}}\ and\ \bibinfo {author} {\bibfnamefont {P.}~\bibnamefont {{Schechter}}},\ }\href {https://doi.org/10.1086/152650} {\bibfield  {journal} {\bibinfo  {journal} {\apj}\ }\textbf {\bibinfo {volume} {187}},\ \bibinfo {pages} {425} (\bibinfo {year} {1974})}\BibitemShut {NoStop}%
\bibitem [{\citenamefont {{Carniani}}\ \emph {et~al.}(2024)\citenamefont {{Carniani}}, \citenamefont {{Hainline}}, \citenamefont {{D'Eugenio}}, \citenamefont {{Eisenstein}}, \citenamefont {{Jakobsen}}, \citenamefont {{Witstok}}, \citenamefont {{Johnson}}, \citenamefont {{Chevallard}}, \citenamefont {{Maiolino}}, \citenamefont {{Helton}}, \citenamefont {{Willott}}, \citenamefont {{Robertson}}, \citenamefont {{Alberts}}, \citenamefont {{Arribas}}, \citenamefont {{Baker}}, \citenamefont {{Bhatawdekar}}, \citenamefont {{Boyett}}, \citenamefont {{Bunker}}, \citenamefont {{Cameron}}, \citenamefont {{Cargile}}, \citenamefont {{Charlot}}, \citenamefont {{Curti}}, \citenamefont {{Curtis-Lake}}, \citenamefont {{Egami}}, \citenamefont {{Giardino}}, \citenamefont {{Isaak}}, \citenamefont {{Ji}}, \citenamefont {{Jones}}, \citenamefont {{Kumari}}, \citenamefont {{Maseda}}, \citenamefont {{Parlanti}}, \citenamefont {{P{\'e}rez-Gonz{\'a}lez}}, \citenamefont {{Rawle}}, \citenamefont {{Rieke}}, \citenamefont {{Rieke}},
  \citenamefont {{Del Pino}}, \citenamefont {{Saxena}}, \citenamefont {{Scholtz}}, \citenamefont {{Smit}}, \citenamefont {{Sun}}, \citenamefont {{Tacchella}}, \citenamefont {{{\"U}bler}}, \citenamefont {{Venturi}}, \citenamefont {{Williams}},\ and\ \citenamefont {{Willmer}}}]{2024Natur.633..318C}%
  \BibitemOpen
  \bibfield  {author} {\bibinfo {author} {\bibfnamefont {S.}~\bibnamefont {{Carniani}}}, \bibinfo {author} {\bibfnamefont {K.}~\bibnamefont {{Hainline}}}, \bibinfo {author} {\bibfnamefont {F.}~\bibnamefont {{D'Eugenio}}}, \bibinfo {author} {\bibfnamefont {D.~J.}\ \bibnamefont {{Eisenstein}}}, \bibinfo {author} {\bibfnamefont {P.}~\bibnamefont {{Jakobsen}}}, \bibinfo {author} {\bibfnamefont {J.}~\bibnamefont {{Witstok}}}, \bibinfo {author} {\bibfnamefont {B.~D.}\ \bibnamefont {{Johnson}}}, \bibinfo {author} {\bibfnamefont {J.}~\bibnamefont {{Chevallard}}}, \bibinfo {author} {\bibfnamefont {R.}~\bibnamefont {{Maiolino}}}, \bibinfo {author} {\bibfnamefont {J.~M.}\ \bibnamefont {{Helton}}}, \bibinfo {author} {\bibfnamefont {C.}~\bibnamefont {{Willott}}}, \bibinfo {author} {\bibfnamefont {B.}~\bibnamefont {{Robertson}}}, \bibinfo {author} {\bibfnamefont {S.}~\bibnamefont {{Alberts}}}, \bibinfo {author} {\bibfnamefont {S.}~\bibnamefont {{Arribas}}}, \bibinfo {author} {\bibfnamefont {W.~M.}\ \bibnamefont {{Baker}}},
  \bibinfo {author} {\bibfnamefont {R.}~\bibnamefont {{Bhatawdekar}}}, \bibinfo {author} {\bibfnamefont {K.}~\bibnamefont {{Boyett}}}, \bibinfo {author} {\bibfnamefont {A.~J.}\ \bibnamefont {{Bunker}}}, \bibinfo {author} {\bibfnamefont {A.~J.}\ \bibnamefont {{Cameron}}}, \bibinfo {author} {\bibfnamefont {P.~A.}\ \bibnamefont {{Cargile}}}, \bibinfo {author} {\bibfnamefont {S.}~\bibnamefont {{Charlot}}}, \bibinfo {author} {\bibfnamefont {M.}~\bibnamefont {{Curti}}}, \bibinfo {author} {\bibfnamefont {E.}~\bibnamefont {{Curtis-Lake}}}, \bibinfo {author} {\bibfnamefont {E.}~\bibnamefont {{Egami}}}, \bibinfo {author} {\bibfnamefont {G.}~\bibnamefont {{Giardino}}}, \bibinfo {author} {\bibfnamefont {K.}~\bibnamefont {{Isaak}}}, \bibinfo {author} {\bibfnamefont {Z.}~\bibnamefont {{Ji}}}, \bibinfo {author} {\bibfnamefont {G.~C.}\ \bibnamefont {{Jones}}}, \bibinfo {author} {\bibfnamefont {N.}~\bibnamefont {{Kumari}}}, \bibinfo {author} {\bibfnamefont {M.~V.}\ \bibnamefont {{Maseda}}}, \bibinfo {author} {\bibfnamefont
  {E.}~\bibnamefont {{Parlanti}}}, \bibinfo {author} {\bibfnamefont {P.~G.}\ \bibnamefont {{P{\'e}rez-Gonz{\'a}lez}}}, \bibinfo {author} {\bibfnamefont {T.}~\bibnamefont {{Rawle}}}, \bibinfo {author} {\bibfnamefont {G.}~\bibnamefont {{Rieke}}}, \bibinfo {author} {\bibfnamefont {M.}~\bibnamefont {{Rieke}}}, \bibinfo {author} {\bibfnamefont {B.~R.}\ \bibnamefont {{Del Pino}}}, \bibinfo {author} {\bibfnamefont {A.}~\bibnamefont {{Saxena}}}, \bibinfo {author} {\bibfnamefont {J.}~\bibnamefont {{Scholtz}}}, \bibinfo {author} {\bibfnamefont {R.}~\bibnamefont {{Smit}}}, \bibinfo {author} {\bibfnamefont {F.}~\bibnamefont {{Sun}}}, \bibinfo {author} {\bibfnamefont {S.}~\bibnamefont {{Tacchella}}}, \bibinfo {author} {\bibfnamefont {H.}~\bibnamefont {{{\"U}bler}}}, \bibinfo {author} {\bibfnamefont {G.}~\bibnamefont {{Venturi}}}, \bibinfo {author} {\bibfnamefont {C.~C.}\ \bibnamefont {{Williams}}},\ and\ \bibinfo {author} {\bibfnamefont {C.~N.~A.}\ \bibnamefont {{Willmer}}},\ }\href
  {https://doi.org/10.1038/s41586-024-07860-9} {\bibfield  {journal} {\bibinfo  {journal} {\nat}\ }\textbf {\bibinfo {volume} {633}},\ \bibinfo {pages} {318} (\bibinfo {year} {2024})},\ \Eprint {https://arxiv.org/abs/2405.18485} {arXiv:2405.18485 [astro-ph.GA]} \BibitemShut {NoStop}%
\bibitem [{\citenamefont {{Abbott}}\ \emph {et~al.}(2018)\citenamefont {{Abbott}} \emph {et~al.}}]{DESY1:3x2pt}%
  \BibitemOpen
  \bibfield  {author} {\bibinfo {author} {\bibfnamefont {T.~M.~C.}\ \bibnamefont {{Abbott}}} \emph {et~al.} (\bibinfo {collaboration} {Dark Energy Survey}),\ }\href {https://doi.org/10.1103/PhysRevD.98.043526} {\bibfield  {journal} {\bibinfo  {journal} {\prd}\ }\textbf {\bibinfo {volume} {98}},\ \bibinfo {eid} {043526} (\bibinfo {year} {2018})},\ \Eprint {https://arxiv.org/abs/1708.01530} {arXiv:1708.01530 [astro-ph.CO]} \BibitemShut {NoStop}%
\bibitem [{\citenamefont {{Beutler}}\ \emph {et~al.}(2011)\citenamefont {{Beutler}}, \citenamefont {{Blake}}, \citenamefont {{Colless}}, \citenamefont {{Jones}}, \citenamefont {{Staveley-Smith}}, \citenamefont {{Campbell}}, \citenamefont {{Parker}}, \citenamefont {{Saunders}},\ and\ \citenamefont {{Watson}}}]{6dF:BAO_H0}%
  \BibitemOpen
  \bibfield  {author} {\bibinfo {author} {\bibfnamefont {F.}~\bibnamefont {{Beutler}}}, \bibinfo {author} {\bibfnamefont {C.}~\bibnamefont {{Blake}}}, \bibinfo {author} {\bibfnamefont {M.}~\bibnamefont {{Colless}}}, \bibinfo {author} {\bibfnamefont {D.~H.}\ \bibnamefont {{Jones}}}, \bibinfo {author} {\bibfnamefont {L.}~\bibnamefont {{Staveley-Smith}}}, \bibinfo {author} {\bibfnamefont {L.}~\bibnamefont {{Campbell}}}, \bibinfo {author} {\bibfnamefont {Q.}~\bibnamefont {{Parker}}}, \bibinfo {author} {\bibfnamefont {W.}~\bibnamefont {{Saunders}}},\ and\ \bibinfo {author} {\bibfnamefont {F.}~\bibnamefont {{Watson}}},\ }\href {https://doi.org/10.1111/j.1365-2966.2011.19250.x} {\bibfield  {journal} {\bibinfo  {journal} {\mnras}\ }\textbf {\bibinfo {volume} {416}},\ \bibinfo {pages} {3017} (\bibinfo {year} {2011})},\ \Eprint {https://arxiv.org/abs/1106.3366} {arXiv:1106.3366 [astro-ph.CO]} \BibitemShut {NoStop}%
\bibitem [{\citenamefont {{Ross}}\ \emph {et~al.}(2015)\citenamefont {{Ross}}, \citenamefont {{Samushia}}, \citenamefont {{Howlett}}, \citenamefont {{Percival}}, \citenamefont {{Burden}},\ and\ \citenamefont {{Manera}}}]{SDSS:DR7}%
  \BibitemOpen
  \bibfield  {author} {\bibinfo {author} {\bibfnamefont {A.~J.}\ \bibnamefont {{Ross}}}, \bibinfo {author} {\bibfnamefont {L.}~\bibnamefont {{Samushia}}}, \bibinfo {author} {\bibfnamefont {C.}~\bibnamefont {{Howlett}}}, \bibinfo {author} {\bibfnamefont {W.~J.}\ \bibnamefont {{Percival}}}, \bibinfo {author} {\bibfnamefont {A.}~\bibnamefont {{Burden}}},\ and\ \bibinfo {author} {\bibfnamefont {M.}~\bibnamefont {{Manera}}} (\bibinfo {collaboration} {BOSS}),\ }\href {https://doi.org/10.1093/mnras/stv154} {\bibfield  {journal} {\bibinfo  {journal} {\mnras}\ }\textbf {\bibinfo {volume} {449}},\ \bibinfo {pages} {835} (\bibinfo {year} {2015})},\ \Eprint {https://arxiv.org/abs/1409.3242} {arXiv:1409.3242 [astro-ph.CO]} \BibitemShut {NoStop}%
\bibitem [{\citenamefont {{Alam}}\ \emph {et~al.}(2017)\citenamefont {{Alam}} \emph {et~al.}}]{SDSS:DR12}%
  \BibitemOpen
  \bibfield  {author} {\bibinfo {author} {\bibfnamefont {S.}~\bibnamefont {{Alam}}} \emph {et~al.} (\bibinfo {collaboration} {BOSS}),\ }\href {https://doi.org/10.1093/mnras/stx721} {\bibfield  {journal} {\bibinfo  {journal} {\mnras}\ }\textbf {\bibinfo {volume} {470}},\ \bibinfo {pages} {2617} (\bibinfo {year} {2017})},\ \Eprint {https://arxiv.org/abs/1607.03155} {arXiv:1607.03155 [astro-ph.CO]} \BibitemShut {NoStop}%
\bibitem [{\citenamefont {{Alam}}\ \emph {et~al.}(2021)\citenamefont {{Alam}} \emph {et~al.}}]{SDSS:DR16}%
  \BibitemOpen
  \bibfield  {author} {\bibinfo {author} {\bibfnamefont {S.}~\bibnamefont {{Alam}}} \emph {et~al.} (\bibinfo {collaboration} {eBOSS}),\ }\href {https://doi.org/10.1103/PhysRevD.103.083533} {\bibfield  {journal} {\bibinfo  {journal} {\prd}\ }\textbf {\bibinfo {volume} {103}},\ \bibinfo {eid} {083533} (\bibinfo {year} {2021})},\ \Eprint {https://arxiv.org/abs/2007.08991} {arXiv:2007.08991 [astro-ph.CO]} \BibitemShut {NoStop}%
\bibitem [{\citenamefont {{Beutler}}\ \emph {et~al.}(2012)\citenamefont {{Beutler}}, \citenamefont {{Blake}}, \citenamefont {{Colless}}, \citenamefont {{Jones}}, \citenamefont {{Staveley-Smith}}, \citenamefont {{Poole}}, \citenamefont {{Campbell}}, \citenamefont {{Parker}}, \citenamefont {{Saunders}},\ and\ \citenamefont {{Watson}}}]{6dF:growth_sigma8}%
  \BibitemOpen
  \bibfield  {author} {\bibinfo {author} {\bibfnamefont {F.}~\bibnamefont {{Beutler}}}, \bibinfo {author} {\bibfnamefont {C.}~\bibnamefont {{Blake}}}, \bibinfo {author} {\bibfnamefont {M.}~\bibnamefont {{Colless}}}, \bibinfo {author} {\bibfnamefont {D.~H.}\ \bibnamefont {{Jones}}}, \bibinfo {author} {\bibfnamefont {L.}~\bibnamefont {{Staveley-Smith}}}, \bibinfo {author} {\bibfnamefont {G.~B.}\ \bibnamefont {{Poole}}}, \bibinfo {author} {\bibfnamefont {L.}~\bibnamefont {{Campbell}}}, \bibinfo {author} {\bibfnamefont {Q.}~\bibnamefont {{Parker}}}, \bibinfo {author} {\bibfnamefont {W.}~\bibnamefont {{Saunders}}},\ and\ \bibinfo {author} {\bibfnamefont {F.}~\bibnamefont {{Watson}}},\ }\href {https://doi.org/10.1111/j.1365-2966.2012.21136.x} {\bibfield  {journal} {\bibinfo  {journal} {\mnras}\ }\textbf {\bibinfo {volume} {423}},\ \bibinfo {pages} {3430} (\bibinfo {year} {2012})},\ \Eprint {https://arxiv.org/abs/1204.4725} {arXiv:1204.4725 [astro-ph.CO]} \BibitemShut {NoStop}%
\bibitem [{\citenamefont {Huterer}\ \emph {et~al.}(2017)\citenamefont {Huterer}, \citenamefont {Shafer}, \citenamefont {Scolnic},\ and\ \citenamefont {Schmidt}}]{Huterer:2016uyq}%
  \BibitemOpen
  \bibfield  {author} {\bibinfo {author} {\bibfnamefont {D.}~\bibnamefont {Huterer}}, \bibinfo {author} {\bibfnamefont {D.}~\bibnamefont {Shafer}}, \bibinfo {author} {\bibfnamefont {D.}~\bibnamefont {Scolnic}},\ and\ \bibinfo {author} {\bibfnamefont {F.}~\bibnamefont {Schmidt}},\ }\href {https://doi.org/10.1088/1475-7516/2017/05/015} {\bibfield  {journal} {\bibinfo  {journal} {JCAP}\ }\textbf {\bibinfo {volume} {05}},\ \bibinfo {pages} {015}},\ \Eprint {https://arxiv.org/abs/1611.09862} {arXiv:1611.09862 [astro-ph.CO]} \BibitemShut {NoStop}%
\bibitem [{\citenamefont {{Said}}\ \emph {et~al.}(2020)\citenamefont {{Said}}, \citenamefont {{Colless}}, \citenamefont {{Magoulas}}, \citenamefont {{Lucey}},\ and\ \citenamefont {{Hudson}}}]{Said:2020}%
  \BibitemOpen
  \bibfield  {author} {\bibinfo {author} {\bibfnamefont {K.}~\bibnamefont {{Said}}}, \bibinfo {author} {\bibfnamefont {M.}~\bibnamefont {{Colless}}}, \bibinfo {author} {\bibfnamefont {C.}~\bibnamefont {{Magoulas}}}, \bibinfo {author} {\bibfnamefont {J.~R.}\ \bibnamefont {{Lucey}}},\ and\ \bibinfo {author} {\bibfnamefont {M.~J.}\ \bibnamefont {{Hudson}}},\ }\href {https://doi.org/10.1093/mnras/staa2032} {\bibfield  {journal} {\bibinfo  {journal} {\mnras}\ }\textbf {\bibinfo {volume} {497}},\ \bibinfo {pages} {1275} (\bibinfo {year} {2020})},\ \Eprint {https://arxiv.org/abs/2007.04993} {arXiv:2007.04993 [astro-ph.CO]} \BibitemShut {NoStop}%
\bibitem [{\citenamefont {{Boruah}}\ \emph {et~al.}(2020)\citenamefont {{Boruah}}, \citenamefont {{Hudson}},\ and\ \citenamefont {{Lavaux}}}]{Boruah:2020}%
  \BibitemOpen
  \bibfield  {author} {\bibinfo {author} {\bibfnamefont {S.~S.}\ \bibnamefont {{Boruah}}}, \bibinfo {author} {\bibfnamefont {M.~J.}\ \bibnamefont {{Hudson}}},\ and\ \bibinfo {author} {\bibfnamefont {G.}~\bibnamefont {{Lavaux}}},\ }\href {https://doi.org/10.1093/mnras/staa2485} {\bibfield  {journal} {\bibinfo  {journal} {\mnras}\ }\textbf {\bibinfo {volume} {498}},\ \bibinfo {pages} {2703} (\bibinfo {year} {2020})},\ \Eprint {https://arxiv.org/abs/1912.09383} {arXiv:1912.09383 [astro-ph.CO]} \BibitemShut {NoStop}%
\bibitem [{\citenamefont {{Turner}}\ \emph {et~al.}(2023)\citenamefont {{Turner}}, \citenamefont {{Blake}},\ and\ \citenamefont {{Ruggeri}}}]{Turner:2023}%
  \BibitemOpen
  \bibfield  {author} {\bibinfo {author} {\bibfnamefont {R.~J.}\ \bibnamefont {{Turner}}}, \bibinfo {author} {\bibfnamefont {C.}~\bibnamefont {{Blake}}},\ and\ \bibinfo {author} {\bibfnamefont {R.}~\bibnamefont {{Ruggeri}}},\ }\href {https://doi.org/10.1093/mnras/stac3256} {\bibfield  {journal} {\bibinfo  {journal} {\mnras}\ }\textbf {\bibinfo {volume} {518}},\ \bibinfo {pages} {2436} (\bibinfo {year} {2023})},\ \Eprint {https://arxiv.org/abs/2207.03707} {arXiv:2207.03707 [astro-ph.CO]} \BibitemShut {NoStop}%
\bibitem [{\citenamefont {Blake}\ \emph {et~al.}(2011)\citenamefont {Blake} \emph {et~al.}}]{Blake:2011}%
  \BibitemOpen
  \bibfield  {author} {\bibinfo {author} {\bibfnamefont {C.}~\bibnamefont {Blake}} \emph {et~al.},\ }\href {https://doi.org/10.1111/j.1365-2966.2011.18903.x} {\bibfield  {journal} {\bibinfo  {journal} {Mon. Not. Roy. Astron. Soc.}\ }\textbf {\bibinfo {volume} {415}},\ \bibinfo {pages} {2876} (\bibinfo {year} {2011})},\ \Eprint {https://arxiv.org/abs/1104.2948} {arXiv:1104.2948 [astro-ph.CO]} \BibitemShut {NoStop}%
\bibitem [{\citenamefont {{Blake}}\ \emph {et~al.}(2013)\citenamefont {{Blake}} \emph {et~al.}}]{Blake:2013}%
  \BibitemOpen
  \bibfield  {author} {\bibinfo {author} {\bibfnamefont {C.}~\bibnamefont {{Blake}}} \emph {et~al.} (\bibinfo {collaboration} {GAMA}),\ }\href {https://doi.org/10.1093/mnras/stt1791} {\bibfield  {journal} {\bibinfo  {journal} {\mnras}\ }\textbf {\bibinfo {volume} {436}},\ \bibinfo {pages} {3089} (\bibinfo {year} {2013})},\ \Eprint {https://arxiv.org/abs/1309.5556} {arXiv:1309.5556 [astro-ph.CO]} \BibitemShut {NoStop}%
\bibitem [{\citenamefont {{Howlett}}\ \emph {et~al.}(2015)\citenamefont {{Howlett}}, \citenamefont {{Ross}}, \citenamefont {{Samushia}}, \citenamefont {{Percival}},\ and\ \citenamefont {{Manera}}}]{Howlett:2015}%
  \BibitemOpen
  \bibfield  {author} {\bibinfo {author} {\bibfnamefont {C.}~\bibnamefont {{Howlett}}}, \bibinfo {author} {\bibfnamefont {A.~J.}\ \bibnamefont {{Ross}}}, \bibinfo {author} {\bibfnamefont {L.}~\bibnamefont {{Samushia}}}, \bibinfo {author} {\bibfnamefont {W.~J.}\ \bibnamefont {{Percival}}},\ and\ \bibinfo {author} {\bibfnamefont {M.}~\bibnamefont {{Manera}}},\ }\href {https://doi.org/10.1093/mnras/stu2693} {\bibfield  {journal} {\bibinfo  {journal} {\mnras}\ }\textbf {\bibinfo {volume} {449}},\ \bibinfo {pages} {848} (\bibinfo {year} {2015})},\ \Eprint {https://arxiv.org/abs/1409.3238} {arXiv:1409.3238 [astro-ph.CO]} \BibitemShut {NoStop}%
\bibitem [{\citenamefont {{Okumura}}\ \emph {et~al.}(2016)\citenamefont {{Okumura}} \emph {et~al.}}]{Okumura:2016}%
  \BibitemOpen
  \bibfield  {author} {\bibinfo {author} {\bibfnamefont {T.}~\bibnamefont {{Okumura}}} \emph {et~al.},\ }\href {https://doi.org/10.1093/pasj/psw029} {\bibfield  {journal} {\bibinfo  {journal} {\pasj}\ }\textbf {\bibinfo {volume} {68}},\ \bibinfo {eid} {38} (\bibinfo {year} {2016})},\ \Eprint {https://arxiv.org/abs/1511.08083} {arXiv:1511.08083 [astro-ph.CO]} \BibitemShut {NoStop}%
\bibitem [{\citenamefont {Pezzotta}\ \emph {et~al.}(2017)\citenamefont {Pezzotta} \emph {et~al.}}]{Pezzotta:2017}%
  \BibitemOpen
  \bibfield  {author} {\bibinfo {author} {\bibfnamefont {A.}~\bibnamefont {Pezzotta}} \emph {et~al.},\ }\href {https://doi.org/10.1051/0004-6361/201630295} {\bibfield  {journal} {\bibinfo  {journal} {Astron. Astrophys.}\ }\textbf {\bibinfo {volume} {604}},\ \bibinfo {pages} {A33} (\bibinfo {year} {2017})},\ \Eprint {https://arxiv.org/abs/1612.05645} {arXiv:1612.05645 [astro-ph.CO]} \BibitemShut {NoStop}%
\bibitem [{\citenamefont {Hirano}\ \emph {et~al.}(2015)\citenamefont {Hirano}, \citenamefont {Zhu}, \citenamefont {Yoshida}, \citenamefont {Spergel},\ and\ \citenamefont {Yorke}}]{Hirano:2015wla}%
  \BibitemOpen
  \bibfield  {author} {\bibinfo {author} {\bibfnamefont {S.}~\bibnamefont {Hirano}}, \bibinfo {author} {\bibfnamefont {N.}~\bibnamefont {Zhu}}, \bibinfo {author} {\bibfnamefont {N.}~\bibnamefont {Yoshida}}, \bibinfo {author} {\bibfnamefont {D.}~\bibnamefont {Spergel}},\ and\ \bibinfo {author} {\bibfnamefont {H.~W.}\ \bibnamefont {Yorke}},\ }\href {https://doi.org/10.1088/0004-637X/814/1/18} {\bibfield  {journal} {\bibinfo  {journal} {Astrophys. J.}\ }\textbf {\bibinfo {volume} {814}},\ \bibinfo {pages} {18} (\bibinfo {year} {2015})},\ \Eprint {https://arxiv.org/abs/1504.05186} {arXiv:1504.05186 [astro-ph.CO]} \BibitemShut {NoStop}%
\bibitem [{\citenamefont {Hirano}\ and\ \citenamefont {Yoshida}(2024)}]{Hirano:2023auh}%
  \BibitemOpen
  \bibfield  {author} {\bibinfo {author} {\bibfnamefont {S.}~\bibnamefont {Hirano}}\ and\ \bibinfo {author} {\bibfnamefont {N.}~\bibnamefont {Yoshida}},\ }\href {https://doi.org/10.3847/1538-4357/ad22e0} {\bibfield  {journal} {\bibinfo  {journal} {Astrophys. J.}\ }\textbf {\bibinfo {volume} {963}},\ \bibinfo {pages} {2} (\bibinfo {year} {2024})},\ \Eprint {https://arxiv.org/abs/2306.11993} {arXiv:2306.11993 [astro-ph.GA]} \BibitemShut {NoStop}%
\bibitem [{\citenamefont {Parashari}\ and\ \citenamefont {Laha}(2023)}]{Parashari:2023cui}%
  \BibitemOpen
  \bibfield  {author} {\bibinfo {author} {\bibfnamefont {P.}~\bibnamefont {Parashari}}\ and\ \bibinfo {author} {\bibfnamefont {R.}~\bibnamefont {Laha}},\ }\href {https://doi.org/10.1093/mnrasl/slad107} {\bibfield  {journal} {\bibinfo  {journal} {Mon. Not. Roy. Astron. Soc.}\ }\textbf {\bibinfo {volume} {526}},\ \bibinfo {pages} {L63} (\bibinfo {year} {2023})},\ \Eprint {https://arxiv.org/abs/2305.00999} {arXiv:2305.00999 [astro-ph.CO]} \BibitemShut {NoStop}%
\bibitem [{\citenamefont {Biagetti}\ \emph {et~al.}(2023)\citenamefont {Biagetti}, \citenamefont {Franciolini},\ and\ \citenamefont {Riotto}}]{Biagetti:2022ode}%
  \BibitemOpen
  \bibfield  {author} {\bibinfo {author} {\bibfnamefont {M.}~\bibnamefont {Biagetti}}, \bibinfo {author} {\bibfnamefont {G.}~\bibnamefont {Franciolini}},\ and\ \bibinfo {author} {\bibfnamefont {A.}~\bibnamefont {Riotto}},\ }\href {https://doi.org/10.3847/1538-4357/acb5ea} {\bibfield  {journal} {\bibinfo  {journal} {Astrophys. J.}\ }\textbf {\bibinfo {volume} {944}},\ \bibinfo {pages} {113} (\bibinfo {year} {2023})},\ \Eprint {https://arxiv.org/abs/2210.04812} {arXiv:2210.04812 [astro-ph.CO]} \BibitemShut {NoStop}%
\end{thebibliography}%
\end{document}